\documentclass[12pt, reqno]{amsart}
\usepackage{amssymb}
\usepackage{amsmath}
\usepackage{amsfonts}
\usepackage{mathrsfs}
\usepackage{graphicx}
\usepackage{placeins}
\usepackage{color}
\usepackage[doublespacing]{setspace}
\usepackage{caption}
\usepackage{subcaption}
\usepackage{natbib}
\usepackage{enumerate}
\usepackage{tabularx}
\usepackage[charter,cal=cmcal]{mathdesign}
\usepackage[colorlinks=true,citecolor=blue,urlcolor=blue,pdfpagemode=UseNone,pdfstartview=FitH]{hyperref}
\usepackage{apptools}
\usepackage{bibunits}
\makeatletter
\def\section{\@startsection{section}{1}
	\z@{1.0\linespacing\@plus\linespacing}{.8\linespacing}{\Large}}

\def\subsection{\@startsection{subsection}{2}
	\z@{.8\linespacing\@plus.7\linespacing}{.7\linespacing}{\large}}

\def\subsubsection{\@startsection{subsubsection}{3}
	\z@{.5\linespacing\@plus.7\linespacing}{-.5em}{\normalfont\bfseries}}
\makeatother

\numberwithin{equation}{section}

\newtheorem{theorem}{Theorem}[section]
\newtheorem{lemma}{Lemma}[section]

\theoremstyle{definition}

\theoremstyle{definition}
\newtheorem{assumption}{Assumption}[section]

\theoremstyle{definition}

\DeclareTextFontCommand{\textbfit}{%
	\fontseries\bfdefault % change series without selecting the font yet
	\itshape
}

\title{}

\begin{document}
	\vspace*{5ex minus 1ex}
	\begin{center}
		\Large \textsc{Counterfactual Analysis under Partial Identification Using Locally Robust Refinement}
		\bigskip
	\end{center}
	
	\date{%
		%TCIMACRO{\TeXButton{Today}{\today}}%
		%BeginExpansion
		\today%
		%EndExpansion
	}
	
	\vspace*{3ex minus 1ex}
	\begin{center}
		Nathan Canen and Kyungchul Song\\
		\textit{University of Houston and University of British Columbia}\\
		\bigskip
		\bigskip
	\end{center}
	
	\thanks{We thank three anonymous referees for their valuable comments. Corresponding author: Kyungchul Song, kysong@mail.ubc.ca, Vancouver School of Economics, University of British Columbia, 6000 Iona Drive, Vancouver, BC, V6t 1Z4, Canada}
	\address{Department of Economics, University of Houston, 3623 Cullen Boulevard - Office 221-C
		Houston, TX, 77204, USA}
	%\email{ncanen@uh.edu}
	\address{Vancouver School of Economics, University of British Columbia, 6000 Iona Drive, Vancouver, BC, V6T 1L4, Canada}
	%\email{kysong@mail.ubc.ca}
	
	\fontsize{12}{14} \selectfont
	
	\begin{abstract}
		Structural models that admit multiple reduced forms, such as game-theoretic models with multiple equilibria, pose challenges in practice, especially when parameters are set-identified and the identified set is large. In such cases, researchers often choose to focus on a particular subset of equilibria for counterfactual analysis, but this choice can be hard to justify. This paper shows that some parameter values can be more ``desirable'' than others for counterfactual analysis, even if they are empirically equivalent given the data. In particular, within the identified set, some counterfactual predictions can exhibit more robustness than others, against local perturbations of the reduced forms (e.g. the equilibrium selection rule). We provide a representation of this subset which can be used to simplify the implementation. We illustrate our message using moment inequality models, and provide an empirical application based on a model with top-coded data. 
		\medskip
		
		\medskip
		
	\end{abstract}
	\maketitle
	
	\begin{bibunit}[econometrica]		
	\section{Introduction}
	
	Economists often use a structural model to analyze the effect of a policy that has never been implemented, such as potential mergers or changes to legislation. For such analysis, it is crucial to have a plausible specification of the causal relationship between the policy variable and the outcome variable of interest. However, a plausible specification can come short of determining the causal relationship uniquely. Prominent examples are found in game theoretic models, where the presence of multiple equilibria admits multiple reduced forms. (\cite{Tamer:03:ReStud} called models with multiple reduced forms \textit{incomplete}.)\footnote{In this paper, a reduced form refers to a system of equations where the variables on the right hand side are all external variables (i.e. inputs) and the variables on the left hand side are all internal variables (i.e. generated through the system of equations using the inputs) in the sense of \cite{Heckman:10:QJE} (p.56). The term ``reduced form'' in this sense appears in econometrics at least as early as in \cite{Koopmans:49:Eca}.} The multiplicity of reduced forms is often the source of set-identification of structural parameters. 
	
	When a parameter is set-identified, all the values in the identified set are empirically equivalent, in the sense that the observed data cannot be used to distinguish between the values. The identified set represents the empirical content of the model for the parameter (i.e., the amount of information on the value of the parameter that can be extracted from data through the model). However, when a parameter is set-identified with a large identified set, the identified set can be of little use for counterfactual analysis in practice. It is not unusual that predictions from counterfactual analysis based on different equilibria yield self-contradictory results (\cite{Aguirregabiria/Nevo:12:WP}, p.111). In this situation, just as in the calibration approach in macroeconomics, the researcher typically uses additional constraints on the parameter values or chooses a specific equilibrium to obtain a meaningful analysis.\footnote{For example, in an entry model, \cite{Jia:08:Eca} focuses on the equilibrium that is most profitable to one of the firms, and  \cite{Ackerberg/Gowrisankaran:06:RAND} use an equilibrium selection rule which randomizes between a Pareto-best equilibrium and a Pareto-worst equilibrium with an unknown probability. Another example is \cite{Roberts/Sweeting:13:AER} in the context of auctions, who consider the particular equilibrium with the lowest signal for agents in the model so they enter the auction.} In other words, the researcher considers a \textit{refinement} of the identified set, by introducing additional restrictions. However, these restrictions are often only weakly motivated.
	
	In our view, a desirable approach in this situation is to \textit{distinguish between the identification step and the refinement step}, and lay out transparently what criterion (or motivation) is used for the refinement step. There is a small body of literature that explores this refinement step. \cite{Song:14:ET} analyzed the problem of choosing a point from the identified interval based on the local asymptotic minimax regret criterion. \cite{Aryal/Kim:13:JBES} considered an incomplete English auction model where the seller is ambiguity averse, and employing the $\Gamma$-maximin expected utility framework of \cite{Gilboa/Schmeidler:89:JME}, suggested a point decision on the optimal reserve price. \cite{Jun/Pinkse:17:WPa} considered an incomplete English auction set-up similar to that in \cite{Haile/Tamer:03:JPE} and proposed a point decision on the optimal reserve price using a maximum entropy method. \cite{Jun/Pinkse:19:JOE} studied a discrete two-player complete information game and established sharp identification of counterfactual predictions of the game. Furthermore, they investigated point decisions on the probability distributions through a maximum entropy method and what they called a Dirichlet approach, and compared the two methods.
	
	The main goal of this paper is to draw attention to the fact that some values in the identified set can give counterfactual predictions that are more robust to the change of certain aspects of model specification than others. Thus, it is reasonable to give priority to those values that produce most ``reliable'' counterfactual predictions. Generally speaking, predictions that are sensitive to a change of the aspects of the model that the researcher is not sure about cannot be reliable. Given that the researcher is often least sure about how a reduced form is determined among multiple reduced forms, we consider a subset of the identified set which produces counterfactual predictions that are more robust to local perturbations of the reduced forms (e.g. local perturbations of the equilibrium selection rule in a game setting) than others.%\footnote{This echoes in spirit \cite{Aguirregabiria/Mira:10:Joe} when they said, ``if the structural model will be used to predict a counterfactual policy, it seems reasonable that the model should be judged in terms of its ability to predict that particular policy. In this sense, the ``best'' model depends on the type of counterfactual policy one wants to predict....'' (p.58).}
	
	More specifically, consider a causal relationship between endogenous variable $y$ and exogenous variables $x,\varepsilon$:
	\begin{align}
	y = \rho_{\beta,\gamma}(x,\varepsilon;\eta),
	\end{align}
	where $\beta$ is the parameter of interest, $\gamma$ a nuisance parameter, and $\eta$ is a random variable with distribution $G$. Without observing $\eta$ or knowing $G$, the model admits multiple reduced forms. For example, in the case of games with multiple equilibria, $G$ plays the role of the equilibrium selection rule.\footnote{As shown in the paper, our approach applies to causally incomplete models other than game-theoretic models, such as models involving top-coded data.} 
	
	Suppose that the counterfactual setting of interest involves a counterfactual distribution $\tilde F_X$ for $X$. Consider the average structural function (ASF) (\cite{Blundell/Powell:04:ReStud}):
	\begin{align*}
		\textsf{ASF}_{\beta,\gamma,G}(x) = \int \int \rho_{\beta,\gamma}(x,\varepsilon;\eta)dG(\eta|x,\varepsilon)dF_\varepsilon(\varepsilon),
	\end{align*}
	where $F_\varepsilon$ denotes the distribution of $\varepsilon$, and $G(\cdot|x,\varepsilon)$ the conditional distribution of $\eta$ given $x$ and $\varepsilon$. Then our proposal is that we choose $\gamma$ that minimizes
	\begin{align}
	\label{LRR0}
	\sup_{G' \in B(G;\delta)}\left| \int \textsf{ASF}_{\beta,\gamma,G}(x)d\tilde F_X(x) - \int \textsf{ASF}_{\beta,\gamma,G'}(x)d\tilde F_X(x)\right|,
	\end{align}
	where $B(G;\delta)$ denotes a $\delta$-neighborhood of $G$ for some small $\delta>0$. (As we show later, given $\beta$, the set of $\gamma$'s that minimize (\ref{LRR0}) does not depend on the choice of $\delta$ or $G$.) We call the resulting set of values $(\beta,\gamma)$ in the identified set \textit{the Locally Robust Refinement (LRR)}. In other words, our refinement is motivated by the desirability of a stable behavior in the counterfactual predictions when the structural parameters are partially identified. 
	
	In many set-ups, the LRR is a substantially smaller set than the identified set. Furthermore, the LRR is often very intuitive. For example, as we show in Section 3, the LRR in an entry game picks nuisance parameter values that minimize the region of multiple equilibria. In the example of a switching regression model with interval data, it picks nuisance parameters that minimize a difference across models in switching regression.
	
	However, unlike the identified set, the size of the LRR does not indicate the empirical content of the model. Rather the LRR gives parameter values which produce counterfactual predictions that are most robust to the change of the reduced forms among their multiple forms. Due to this robustness property, one may want to pay more attention to those parameter values than others when conducting policy exercises, even if all the values in the identified set are observationally equivalent. 
	
	The formulation in (\ref{LRR0}) gives the impression that the LRR involves  additional optimization over an infinite dimensional space in estimation, adding to computational cost. Using the linearity of $\textsf{ASF}_{\beta,\gamma,G}(x)$ in $G$ and Hilbert space geometry, this paper gives a simple characterization of the LRR so that focusing on the LRR does not cause extra computational cost. In fact, due to the additional constraints that come from the LRR, the computational cost is substantially reduced from a benchmark case without using the LRR. Focusing on moment inequality models, this paper presents a method of bootstrap inference on the LRR from the identified set and shows its uniform asymptotic validity. We provide results from Monte Carlo simulations, and illustrate the LRR through an empirical application on top-coded data. These exercises illustrate that the LRR's predictions are indeed more stable than those with the identified set and that they may affect policymaking in applied contexts (i.e. policies suggested by the identified set could be driven by parameter values that are not robust, and are not part of the LRR).
	
	There has long been interest in studying model sensitivity and robustness to misspecifications in econometric models. Recent contributions in this line include \cite{Andrews/Gentzkow/Shapiro:18:NBER}, \cite{Armstrong/Kolesar:20:QE} and \cite{Bonhomme/Weidner:20:WP} who, among many others, proposed a method of measuring local misspecification and performing inference that is robust to local misspecifications. In the context of instrumental variable models, \cite{Nevo/Rosen:12:ReStat} permitted the instrumental variable to be correlated with the error term and presented bounds for the parameter of interest under an assumption on the correlation between the  instrumental variable and the error term. \cite{Oster:19:JBES} proposed a method of extracting information on the omitted variable bias from data in linear regression models. \cite{Masten/Poirier:18:Ecma} provided an identification analysis of various treatment effect parameters from treatment effect models where the conditional independence restriction is weakened to a partial dependence condition.\footnote{Robustness to misspecification has also been considered in the macroeconomics literature (e.g. in optimal control theory as in \cite{Hansen/Sargent:01:AER}, see \cite{Hansen/Marinacci:16:SS} as well), where one can consider perturbations of the decision maker's model, but less emphasis is placed on the identification and inference problems in these models. Finally, \cite{Andrews/Gentzkow/Shapiro:17:QJE} focus on the case where the model is point-identified, and present a statistic on how estimates vary when a model's moments changes.} \cite{Christensen/Connault:19:WP} proposed a computationally tractable method of performing inference on counterfactual quantities that arise in structural models, while permitting a global misspecification of the model. \cite{Li:19:WP} uses a semiparametric restriction (instead of parametric ones) to generate bounds for counterfactuals of interest in moment inequality models.
	
	All these authors are concerned about robustness of inference to misspecifications. Their robust inference often leads to a larger confidence set, and a wider range of policy predictions. In contrast, we consider settings where the identified set is already too large to be useful for a meaningful policy analysis. This can happen \textit{even if} the model is correctly specified (see \cite{Aguirregabiria/Nevo:12:WP} for a discussion, particularly in the context of multiple equilibria). Instead of enlarging an existing identified set by ``robustifying'' it as above, we explore a rationale for focusing on a subset of the identified set for policy analysis. Our criterion is based on the stability of counterfactual predictions as we perturb the reduced-form selection rules. One can relate our guiding philosophy to the literature of equilibrium refinements in game theory, where one focuses on a subset of the set of Nash equilibria based on its stability against various perturbations of the game. (See Chapter 5 of \cite{Myerson:91:GameTheory}.) Then the idea of local robustness in the paper, i.e., the stability of reduced forms against local perturbations of reduced form selection rules is related to \cite{Fudenberg/Kreps/Levine:88:JET} who considered robustification of the set of strict Nash equilibria against local perturbations of the games.
	
	The paper is organized as follows. The next section gives a general set-up of a structural model and presents examples that fit our framework. Section 3 presents the approach of LRR and provides its characterization that is useful for implementation in practice. Focusing on generic moment inequality models, Section 4 proposes  inference for the LRR and presents results from Monte Carlo simulations. An empirical application is presented in Section 5. Section 6 concludes. In the Appendix, we provide some mathematical proofs. The Supplemental Note to this paper includes additional information on the data for the empirical application, and presents results on the uniform validity of our inference procedure.
	
	\section{A Counterfactual Analysis with Multiple Reduced Forms}
	\subsection{Models with Multiple Reduced Forms}
	Suppose that data $\{(Y_i',X_i')'\}_{i=1}^n$ are drawn from the joint distribution $P$ of $(Y',X')'$ which obeys the following \textit{reduced form}: for $Y \in \mathbf{R}$ and $W = (X',\varepsilon')'$,
	\begin{align}
	\label{true model}
	Y = \rho(W;\eta),
	\end{align}
	for some function $\rho$, where $X$ is observed but $\varepsilon$ and $\eta$ are not. For each value $\eta$, we regard the map $\rho(\cdot;\eta)$ as representing a functional-causal relationship between internal variable $Y$ and external variables $W$.\footnote{Here we follow the proposal by \cite{Heckman:10:QJE} (p.56) and use the terms \textit{internal variables} and \textit{external variables}. External variables are specified from outside of the structural equations as inputs, and internal variables are generated through the system of equations from the inputs. External variables are not necessarily exogenous variables. When $X$ is correlated with $\varepsilon$, the variable $X$ is called endogenous, although $(X,\varepsilon)$ is jointly given externally as inputs, and thus is a vector of external variables. Here a reduced form refers to a system of equations where all the variables on the left hand side are internal variables and all the variables on the right hand side are external variables.} This relationship is fully determined once the value of $\eta$ is realized. Typically, $\varepsilon$ has a structural meaning such as unobserved costs, whereas $\eta$ does not and is used only as an index for a reduced form. For example, in a game-theoretic model, $W = (X',\varepsilon')'$ represents observed and unobserved payoff components and $\eta$ an index for an equilibrium. The response of $Y$ to a change in $W$ is ambiguous unless the value of $\eta$ is fixed. In this sense, there is a multiplicity of reduced forms in (\ref{true model}). To describe the generation of observed outcome $Y$, we assume that Nature selects the value of $\eta$ from the conditional distribution $G(\cdot|w)$ given $W = w$. In a game setting, $G$ is often called the equilibrium selection rule. However, the multiplicity of reduced forms can arise in a setting that has nothing to do with a game, such as in the case of switching regressions with interval data which we describe below. Hence we call $G$ the \textit{reduced-form selection rule} in this paper.
	
	\subsection{Counterfactual Analysis Under Partial Identification}
	
	We focus on a counterfactual experiment where the distribution $F$ of the external vector $W$ is changed to a counterfactual one $\tilde F$. Such a counterfactual experiment includes cases where one ``shuts off'' the effect of a certain covariate, ``fixing'' of the covariate to a certain value, or changing the distribution of $\varepsilon$. Suppose that the counterfactual quantity of interest is generically denoted by $C(\rho,\tilde F,G)$. This quantity could be, for example, the prediction of game outcomes or expected payoffs under a counterfactual scenario:
	\begin{align*}
		C(\rho,\tilde F,G) = \int \rho(w;\eta)dG(\eta|w) d\tilde F(w). 
	\end{align*}
	Or in other applications, $C(\rho,\tilde F,G)$ could be the average increase in welfare. Let $\mathcal{G}$ be the space of the reduced-form selection rules $G$ and let $\mathcal{R}$ be the identified set for the function $\rho$. Then the identified set for the counterfactual prediction $C(\rho,\tilde F,G)$ is given by
	\begin{align*}
		\{C(\rho,\tilde F,G):G \in \mathcal{G}, \rho \in \mathcal{R}\}.
	\end{align*}
	A standard approach to make inference based on this identified set faces multiple challenges in practice. The set of predictions can be of little practical use, when the estimated set is too large. Furthermore, estimation of the identified set itself and statistical inference can be computationally intensive.
	
	Due to these difficulties, it is not unusual that empirical researchers focus on a particular subset of $\mathcal{G}$ (such as focusing on a particular equilibrium) or consider a subset of $\mathcal{R}$ (such as by restricting the admissible values of parameters as in the case of calibration). Such a subset selection approach, while inevitable in practice, is not easy to motivate.
	
	\subsection{Examples}
	
	For brevity, we consider two examples here to illustrate our set-up: a finite game of complete information and a switching regression with interval data.
	
	\subsubsection{Finite Game of Complete Information} \label{sec: entry game complete info}
	Consider the setting of a complete information $N$ player game with finite actions, and let $A$ be the finite action space for each player.\footnote{A special case is the two player entry game used by \cite{Bresnahan/Reiss:91:JOE} and \cite{Tamer:03:ReStud}, among others.} Suppose that the payoff for player $i$ playing $\tilde a \in A$ when all other players are playing $a_{-i} = (a_j)_{j \ne i}$ is given by $u_i(\tilde a, a_{-i},W)$ for a map $u_i$, where $W = (X,\varepsilon)$ is a payoff-relevant  characteristic vector of player $i$. For each $a \in A^N$, we define
	\begin{align}
	\mathbb{W}(a) = \left\{w: u_i(a_i,a_{-i},w) \ge \max_{a' \in A} u_i(a',a_{-i},w), \forall i=1,...,N \right\}.
	\end{align}
	Hence if $w$ belongs to $\mathbb{W}(a)$, $a$ is a Nash equilibrium for the game with $w$, so that the set
	\begin{align}
	\mathcal{A}(w) = \left\{a \in A^N: w \in \mathbb{W}(a) \right\}
	\end{align} 
	represents the set of equilibria for the game with $w$. Let $\eta$ be a random vector taking values in $A^N$ where the conditional distribution $G(\cdot|w)$ of $\eta$ given $w$ has a support in $\mathcal{A}(w)$. Let $Y = (Y_i)_{i =1 }^N$ be the observed action profile that comes from a Nash equilibrium $\eta$ of the game after Nature draws $\eta$ from $G(\cdot|w)$. Then, we can write for $i \in N$,
	\begin{align}
	Y_i = \rho_i(W;\eta),
	\end{align}
	where $\rho_i$ is the $i$-th entry of $\rho$ which is defined as
	\begin{align*}
		\rho(w;\eta) = \sum_{a \in \mathcal{A}(w)} 1\{\eta = a\} a.
	\end{align*}
	The conditional distribution $G(\cdot|w)$ over $\eta$ represents the equilibrium selection rule.
	
	\subsubsection{A Switching Regression Model with Interval Outcomes\label{interval_data_ex}}
	
	Consider an interval data set-up similar to one studied by \cite{Manski/Tamer:02:ECMA}. More specifically, we consider top-coded data with threshold variables $Z_1 < Z_2$, where $Z_1, Z_2$ are observed and common across individuals. We observe an outcome random variable which equals a latent $Y$ if $Y  \leq Z_1$ and is only known to be within $[Z_1, Z_2]$ otherwise, and a vector of covariates $X_1$. For example, $X_1$ represents a vector of covariates for each worker such as age, education, experience, gender, and $Y$ is the log of the hourly wage.
	
	It is assumed that the latent $Y$ follows:
	\begin{eqnarray*}
		Y = \left\{\begin{array}{ll}
			      g_1\left(\beta_0 D + X_1'\gamma_0 \right) + \varepsilon, &\text{ if } \eta = 1, \text{ and }\\
			      g_2\left(\beta_0 D + X_1'\gamma_0 \right) + \varepsilon, & \text{ if } \eta = 0,
			    \end{array}
		        \right.
	\end{eqnarray*}
	where $D$ denotes the policy variable, and $\eta$ the binary indicator of a regime, and $g_1$ and $g_2$ are some known functions. As for the error term $\varepsilon$, we assume that $\mathbf{E}[\varepsilon|X_1,D] = 0$. We also assume that $Y \le Z_2$ with probability one. Let $\tilde Y_1 = \tilde Y_2 = Y$ if $Y \le Z_1$, and $\tilde Y_1 = Z_1$ and $\tilde Y_2 = Z_2$ if $Y > Z_1$. The econometrician observes $\tilde Y_1$ and $\tilde Y_2$. Let $\pi_0(X_1,D) = P\{\eta = 1 |X_1,D\}$ and $\Pi$ be the collection of functions that $\pi_0$ is known to belongs to. 
	
	It is straightforward that the identified set for $(\beta_0,\gamma_0)$ is given by
	\begin{equation}
	\label{idset_interval}
	\Theta_P = \left\{(\beta_0,\gamma_0): \mathbf{E}[\tilde Y_1 |X_1,D] \leq \alpha(X_1,D;\beta_0,\gamma_0,\pi) \leq \mathbf{E}[\tilde Y_2 |X_1, D], \text{a.e.} \text{ for some } \pi \in \Pi \right\},
	\end{equation}
    where
    \begin{align}
    	\label{alpha}
    	\alpha(X_1,D;\beta_0,\gamma_0,\pi) =  g_1\left(\beta_0 D + X_1'\gamma_0 \right)  \pi(X_1,D) + g_2\left(\beta_0 D + X_1'\gamma_0 \right)  (1 - \pi(X_1,D)).
    \end{align}
     Now we can write
	\begin{align*}
		Y = \rho(X,\varepsilon;\eta),
	\end{align*}
	where $X = (X_1,D)$, and 
	\begin{align}
	\label{g beta gamma Interval}
	\rho(X,\varepsilon;\eta)  = g_1\left(\beta_0 D + X_1'\gamma_0 \right) \eta + g_2\left(\beta_0 D + X_1'\gamma_0 \right) (1 - \eta) + \varepsilon.
	\end{align}
	The distribution $G$ of $\eta$ plays a role analogous to the equilibrium selection rule in a game-theoretic model. Here it selects from a given set of reduced forms.
	
	\section{Locally Robust Refinement}
	\subsection{Overview\label{overview_section}}
	Suppose that the reduced-form $\rho(\cdot;\eta)$ in (\ref{true model}) is parametrized as follows:
	\begin{align}
	Y = \rho_{\beta_0,\gamma_0}(W;\eta),
	\end{align}
	where $\beta_0$ is a parameter of interest taking values in a set $B$ and $\gamma_0$ is a nuisance parameter taking values in a set $\Gamma$, and $\eta$ belongs to some known compact set denoted by $H$. Let $\Theta_P$ be the identified set for $(\beta_0,\gamma_0)$. The counterfactual experiment involves changing the distribution $F$ of $W$ to a counterfactual distribution $\tilde F$. Let us consider the counterfactual average structural outcome:
	\begin{align}
	\label{ASO}
	\textsf{ASO}_{\beta_0,\gamma_0}(G) = \int  \int \rho_{\beta_0,\gamma_0}(w;\eta)dG(\eta|w) d\tilde F(w).
	\end{align}
	This is the average value of outcome $Y$ when a counterfactual change is made to the distribution of $W$ while everything else remains the same. Our focus is on a subset of $\Theta_P$ at which $\textsf{ASO}_{\beta, \gamma}(G)$ behaves robustly as we perturb the reduced-form selection rule $G$.  
		
	More formally, for any $K >0$, we introduce a $K$-sensitivity of $\textsf{ASO}_{\beta,\gamma}(G)$ as we perturb $G$ around in its $K$-neighborhood:
	\begin{align}
		\label{S beta gamma}
		S_{\beta,\gamma}(G;K) = \sup_{G' \in \mathcal{G}: \delta_{\mathcal{G}}(G',G) \le K} \frac{|\textsf{ASO}_{\beta,\gamma}(G') - \textsf{ASO}_{\beta,\gamma}(G)|}{\delta_{\mathcal{G}}(G',G)},
	\end{align}
    where $\mathcal{G}$ denotes the collection of $G$'s permitted in the model. The quantity $S_{\beta,\gamma}(G;K)$ measures the maximal rate of change of counterfactual average outcome when $G$ is perturbed within its $K$-neighborhood, where this neighborhood is defined using a metric $\delta_{\mathcal{G}}(G',G)$, formally defined below. We search for values of nuisance parameters $\gamma$ which minimize this sensitivity.  For each $\beta \in B$, if we let
	\begin{align}
		\label{Gamma LRR}
		\Gamma^{\mathsf{LRR}}(\beta) = \text{argmin}_{\gamma: (\beta,\gamma) \in \Theta_P} S_{\beta,\gamma}(G;K),
	\end{align}
	we define
	\begin{align}
	\label{LRR4}
	\Theta_P^{\mathsf{LRR}}  = \{(\beta,\gamma) \in \Theta_P: \gamma \in \Gamma^{\mathsf{LRR}}(\beta) \}.
	\end{align}
	
	We call the subset $\Theta_P^{\mathsf{LRR}}$ the \textit{Locally Robust Refinement (LRR)} of $\Theta_P$. Hence, the LRR can be viewed as a collection of parameters at which the ASO is stable against local misspecifications of $G$.	As we will see later in the characterization of the LRR set, the set is independent of the choice of $K$, $G$ and $\mathcal{G}$. The main reason behind this independence is that the functional $\textsf{ASO}_{\beta,\gamma}(G)$ is linear in $G$. (This is analogous to that the slope of a linear function is constant on the domain of the function.)\footnote{As long as the true reduced-form selection rule lies in the interior of $\mathcal{G}$, our characterization of the LRR set does not depends on the choice of $\mathcal{G}$. While it is possible to recover some information about the rule $G$ with further specifications, our procedure remains agnostic about $\mathcal{G}$, which is in line with many existing works in the econometrics literature of game-theoretic models - see \cite{Ciliberto/Tamer:09:Eca} for one such example.} Note that, as we do not want our concern of local robustification against the reduced-form selection rules to directly affect the parameter of interest, our approach subjects only the nuisance parameters to the refinement. 
	
	The inference on $\beta$ based on the Locally Robust Refinement proceeds under the further restriction as follows.
	\begin{align}
	\label{LRR Hypothesis}
	(\beta_0,\gamma_0) \in \Theta_P^{\mathsf{LRR}}.
	\end{align}
	We call this restriction \textit{the LRR Hypothesis}. Once we focus on the LRR set of parameters, this can be used for various exercises of counterfactual analysis, just as one does with the identified set of parameters. 		
	Despite the impression that finding the set $\Theta_P^{\mathsf{LRR}}$ might be computationally intensive as it involves layers of optimization, it is not, due to our characterization result in the next section.
	
	\subsection{Characterization of the Locally Robust Refinement}
	To derive a characterization of the LRR, let us introduce some geometric structure on the space of reduced-form selection rules $G$. Let $\mathcal{G}$ be the collection of all the conditional distributions $G$ (of $\eta$ given $W$) which are dominated by the uniform distribution $\mu$ over $H$, and $w \in \mathcal{W}$, with $\mathcal{W}$ denoting the set $W$ takes values from. Let us introduce the following metric on $\mathcal{G}$: for $G,G' \in \mathcal{G}$,
	\begin{align}
	\delta_\mathcal{G}(G',G) = \sqrt{\int \int \left( \frac{dG'}{d\mu} - \frac{dG}{d\mu}\right)^2(\eta|w) d\mu(\eta) d\tilde F_P(w)}.
	\end{align}
	
	Let us define
	\begin{align}
	\label{Gamma P beta}
		\Gamma(\beta) = \{\gamma \in \Gamma: (\beta,\gamma) \in \Theta_P\},
	\end{align}
	i.e., the set of nuisance parameters that are admitted in the counterfactual analysis together with given $\beta$. For $\kappa\ge0$, we define the $\kappa$-LRR set for $\gamma$ corresponding to $\beta$ as
	\begin{align}
	\quad \quad
	\Gamma_{\kappa}^{\mathsf{LRR}}(\beta) = \left\{\gamma \in \Gamma(\beta): S_{\beta,\gamma}^2(G;K) \le S_{\beta,\tilde \gamma}^2(G;K)+\kappa, \forall \tilde \gamma \in \Gamma(\beta) \right\},
	\end{align}
    where $S_{\beta,\gamma}(G;K)$ is defined in (\ref{S beta gamma}). The set $\Gamma^{\mathsf{LRR}}(\beta)$ defined in (\ref{Gamma LRR}) is a special case of $\Gamma_{\kappa}^{\mathsf{LRR}}(\beta)$ with $\kappa = 0$. However, as we discuss in Section 4.1, it is convenient to use the LRR set with a small positive number $\kappa>0$ which allows us to develop uniform inference. As shown below, this set does not depend on the choice of $K>0$ or that of $G \in \mathcal{G}$.
	
	We now provide a useful characterization of the set $\Gamma_{\kappa}^{\mathsf{LRR}}(\beta)$. For this, we define
	\begin{align*}
		\Delta_{\beta,\gamma}(w;\eta) =  \rho_{\beta,\gamma}(w;\eta) - \int  \rho_{\beta,\gamma}(w;\eta) d\mu(\eta).
	\end{align*}
	The function $\Delta_{\beta,\gamma}(w;\eta)$ is a mean-deviation form of the structural function $\rho_{\beta,\gamma}(w;\eta)$. This function plays a central role in the characterization result. Let us make the following assumption that guarantees that the average variability of this mean-deviation is well defined.
	
	\begin{assumption}
		\label{assump: L2 bound CASF}
		For each $\beta \in B$ and $\gamma \in \Gamma(\beta)$, $Q^\mathsf{LRR}(\beta,\gamma) < \infty$, where
		\begin{align}
		\label{Q LRR}
		Q^\mathsf{LRR}(\beta,\gamma) = \int \int \Delta_{\beta,\gamma}^2(w;\eta) d\mu(\eta) d\tilde F(w).
		\end{align}
	\end{assumption}
	In applications where we can compute $\rho_{\beta,\gamma}(w;\eta)$ for each value of $\eta$, we can evaluate $Q^\mathsf{LRR}(\beta,\gamma)$ easily. (See Section 3.3 below for examples.) The following theorem shows that the set $\Gamma_{\kappa}^{\mathsf{LRR}}(\beta)$ is fully characterized through the function $Q^\mathsf{LRR}(\beta,\gamma)$.
	
	\begin{theorem}
		\label{thm: characterization}
		Suppose that Assumption \ref{assump: L2 bound CASF} holds. Then for each $\beta \in B$ and $\gamma \in \Gamma(\beta)$,
		\begin{align}
		S_{\beta,\gamma}(G;K) = \sqrt{Q^\mathsf{LRR}(\beta,\gamma)},
		\end{align}
		for all $G \in \mathcal{G}$ and all $K>0$.
	\end{theorem} 
	
	Thus the $\kappa$-LRR of $\gamma_0$ can be rewritten as
	\begin{align}
	\label{LRR3}
	\Gamma_{\kappa}^{\mathsf{LRR}}(\beta) = \left\{\gamma \in \Gamma(\beta): Q^\mathsf{LRR}(\beta,\gamma) \le Q^\mathsf{LRR}(\beta,\bar \gamma) + \kappa, \text{ for all } \bar \gamma \in \Gamma(\beta) \right\}.
	\end{align}
	The $\kappa$-LRR of the nuisance parameter vector $\gamma_0$ is the set of $\gamma$'s which minimize the average variability of $\rho_{\beta,\gamma}(w;\eta)$ (around its mean) as we perturb $\eta$. Thus, the $\kappa$-LRR for $(\beta_0,\gamma_0)$ is given by
	\begin{align}
	\label{B_LRR}
		\Theta_{\kappa,P}^{\mathsf{LRR}} = \left\{(\beta,\gamma) \in \Theta_P: \gamma \in \Gamma_{\kappa}^{\mathsf{LRR}}(\beta) \right\}.
	\end{align}

   Sometimes, the target parameter may not be a parameter $\beta$ that determines the reduced form $\rho_{\beta,\gamma}$ but some aggregate quantity such as Average Treatment Effects or aggregate welfare. Suppose that such a target parameter is identified up to $\beta_0$ and $\gamma_0$, so that we can write it as, say $\psi(\beta_0,\gamma_0)$. (We give an example in Section \ref{ex_interval} below.) Then, the LRR set for $\psi(\beta_0,\gamma_0)$ is given by
   \begin{align}
   	   \Psi_{\kappa}^{\mathsf{LRR}} = \left\{ \psi(\beta,\gamma): (\beta,\gamma) \in \Theta_{\kappa,P}^{\mathsf{LRR}}\right\}.
   \end{align}
   To build familiarity with our characterization results for the LRR and show how it can be used in practice, we now derive $Q^{LRR}$ for two applied examples.
	
	\subsection{Examples}
	\subsubsection{Entry Game with Nash Equilibria}
\begin{figure}[h!]
\begin{center}
	\includegraphics[scale=0.37, trim = 0cm 2.7cm 0cm 1.5cm, clip]{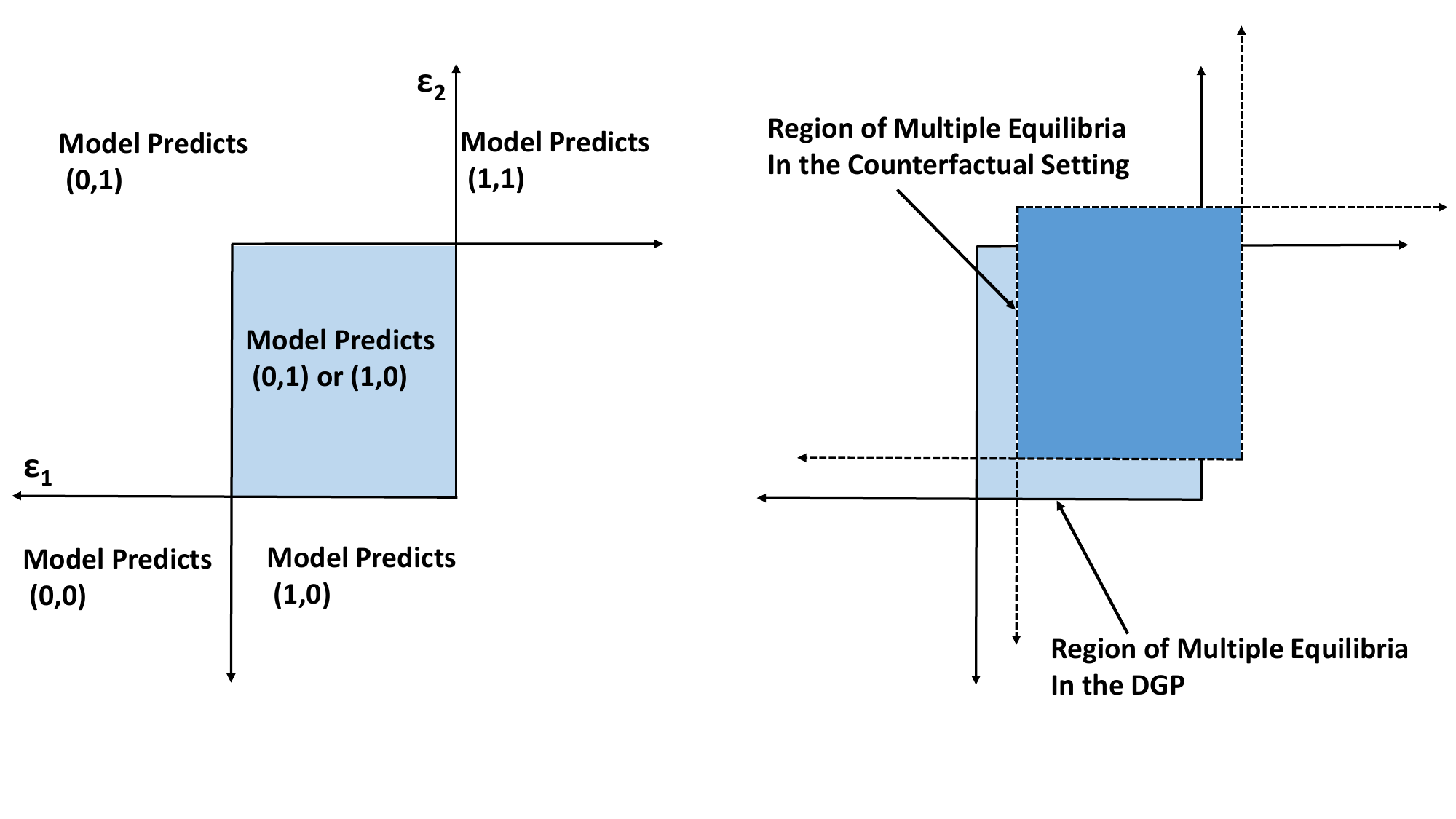}
	\caption{Locally Robust Refinement for the Entry Game Example}\medskip
	\parbox{6.2in}{\footnotesize
		Notes: The figure is based on Figure 1 of \cite{Ciliberto/Tamer:09:Eca} and presents a graphical analysis of the LRR for the entry game example. In the data, the region in light blue is the region of multiple equilibria which is the set $A_{3,\beta,\gamma}(X) \cap A_{4,\beta,\gamma}(X)$. In our counterfactual setting involving $X$ changed to, say, $\tilde X$, this region becomes $A_{3,\beta,\gamma}(\tilde X) \cap A_{4,\beta,\gamma}(\tilde X)$ depicted in dark blue in the right hand side figure. The LRR chooses nuisance parameters in their identified set that minimize the probability mass in the dark blue region. \bigskip}		
	\label{fig: entry game}
\end{center}
\end{figure}

	Consider the setting of a 2x2 complete information entry games, such as the empirical setting in \cite{Bresnahan/Reiss:91:JOE} and \cite{Ciliberto/Tamer:09:Eca}, where there are two firms, $i=1,2$, deciding whether to enter a market or not. The entry decisions of the firms can then be formulated as
	\begin{align}
		\label{entry_system}
		D_1 &= 1\{\beta_{1,0} D_2 + X'\gamma_{1,0} \geq \varepsilon_1\}, \text{ and } \\
		\label{entry_system2}
		D_2 &= 1\{\beta_{2,0} D_1 + X'\gamma_{2,0} \geq \varepsilon_2\},
	\end{align}
	where $\beta_{1,0} D_2 + X'\gamma_{1,0} - \varepsilon_1$ represents the payoff difference for firm $1$ between entering and not entering the market, $D_i = 1$ represents the entry decision by firm $i$ and $D_i = 0$ the decision not to enter by firm $i$. The random vectors $X$ and $\varepsilon$ represent respectively the observed and unobserved characteristics of firms $1$ and $2$ combined. The coefficients $\beta_{1,0}$ and $\beta_{2,0}$ capture strategic interactions between firms and they are our coefficients of interest, whereas the coefficients $\gamma_{1,0}$ and $\gamma_{2,0}$ measure the roles of firm specific and market specific characteristics and, as such, are nuisance parameters.
	\begin{center}
		[INSERT FIGURE \ref{fig: entry game} HERE]
	\end{center}
	In order to find a reduced form as in (\ref{true model}), we first put 
	\begin{align*}
		Y = (D_1,D_2),\text{ and } \varepsilon = (\varepsilon_1,\varepsilon_2),
	\end{align*}
	where $Y \in \mathcal{Y} = \{(0,0),(1,1),(0,1),(1,0)\}$. Define (with $\beta_0 = (\beta_{1,0}',\beta_{2,0}')'$ and $\gamma_0 = (\gamma_{1,0}',\gamma_{2,0}')'$)
	\begin{align*}
		A_{1,\beta_0,\gamma_0}(X) &= \{(e_1,e_2) \in \mathbf{R}^2: X'\gamma_{1,0} < e_1, X'\gamma_{2,0} < e_2\}\\
		A_{2,\beta_0,\gamma_0}(X) &= \{(e_1,e_2) \in \mathbf{R}^2: \beta_{1,0} + X'\gamma_{1,0} \ge e_1, \beta_{2,0} + X'\gamma_{2,0} \ge e_2\}\\
		A_{3,\beta_0,\gamma_0}(X) &= \{(e_1,e_2) \in \mathbf{R}^2: \beta_{1,0} + X'\gamma_{1,0} < e_1, X'\gamma_{2,0} \ge e_2\}, \text{ and }\\
		A_{4,\beta_0,\gamma_0}(X) &= \{(e_1,e_2) \in \mathbf{R}^2: X'\gamma_{1,0} \ge e_1, \beta_{2,0} + X'\gamma_{2,0} < e_2\}.
	\end{align*}
	These regions are represented on the left panel of Figure \ref{fig: entry game}.
	
	Let $\rho_{\beta,\gamma}(x,\varepsilon;\eta) = (\rho_{\beta,\gamma,1}(x,\varepsilon;\eta),\rho_{\beta,\gamma,2}(x,\varepsilon;\eta))$, where for a random variable $\eta \in \{0,1\}$, 
	\begin{align}
		\label{asf game1}
		\quad \rho_{\beta,\gamma,1}(x,\varepsilon;\eta) &= 1\{(\varepsilon_1,\varepsilon_2) \in A_{2,\beta,\gamma}(x)\} +  1\{(\varepsilon_1,\varepsilon_2) \in A_{4 \setminus 3,\beta,\gamma}(x)\}  \\
		& \quad +  1\{(\varepsilon_1,\varepsilon_2) \in A_{3 \cap 4,\beta,\gamma}(x)\} \eta \notag \\
		\rho_{\beta,\gamma,2}(x,\varepsilon;\eta) &= 1\{(\varepsilon_1,\varepsilon_2) \in A_{2,\beta,\gamma}(x)\} + 1\{(\varepsilon_1,\varepsilon_2) \in A_{3 \setminus 4,\beta,\gamma}(x)\} \\
		&\quad +  1\{(\varepsilon_1,\varepsilon_2) \in A_{3 \cap 4,\beta,\gamma}(x)\} \eta \notag,
	\end{align}
where $A_{3 \cap 4,\beta,\gamma}(x) = A_{3,\beta,\gamma}(x) \cap A_{4,\beta,\gamma}(x)$, $A_{4 \setminus 3,\beta,\gamma}(x) = A_{4,\beta_0,\gamma_0}(x) \setminus A_{3,\beta_0,\gamma_0}(x)$, and $A_{3 \setminus 4,\beta,\gamma}(x) = A_{3,\beta_0,\gamma_0}(x) \setminus A_{4,\beta_0,\gamma_0}(x)$.
	Then we can write reduced forms for $Y_i$, for firm $i = 1,2$ as
	\begin{align*}
		Y_i = \rho_{\beta_0,\gamma_0,i}(X,\varepsilon;\eta).
	\end{align*}
		Let the distribution $G$ of $\eta$ be a discrete distribution with support $\{0,1\}$, and take $\mu$ to be a Bernoulli random variable with a mean $1/2$.\footnote{This example focuses on pure-strategy Nash equilibria following the empirical literature on entry-exit games (e.g. \cite{Tamer:03:ReStud}, \cite{Jia:08:Eca}, \cite{Ciliberto/Tamer:09:Eca}, among many others).} It follows that for $i = 1,2$,
	\begin{align}
		\label{eq43}
		\rho_{\beta,\gamma,i}(x,\varepsilon;\eta)-\int \rho_{\beta,\gamma,i}(x,\varepsilon;\eta) d\mu(\eta) 
		= \left(\eta-\frac{1}{2}\right) 1\{(\varepsilon_1,\varepsilon_2) \in A_{3 \cap 4,\beta,\gamma}(x)\}.
	\end{align}
   Hence, by Theorem \ref{thm: characterization}, the LRR set based on the ASO of $Y_1$ is the same as that based on the ASO of $Y_2$. Writing the mean-deviated reduced form on the left hand side of (\ref{eq43}) as $\Delta_{\beta,\gamma}(x,\varepsilon;\eta)$, we find that
	\begin{align*}
		\Delta_{\beta,\gamma}^2(x,\varepsilon;\eta) = \left(\eta-\frac{1}{2}\right)^2 1\{(\varepsilon_1,\varepsilon_2) \in A_{3 \cap 4,\beta,\gamma}(x)\}.
	\end{align*}
	
	Plugging this into (\ref{Q LRR}), we find that
	\begin{align}
		\label{Q LRR prelim}
		Q^\mathsf{LRR}(\beta,\gamma) &= \int \int \left(\eta-\frac{1}{2}\right)^2 d\mu(\eta) 1\{(\varepsilon_1,\varepsilon_2) \in A_{3 \cap 4,\beta,\gamma}(x)\} d\tilde F_{X,\varepsilon}(x,\varepsilon) \notag \\ 
		&= \frac{1}{4}\int 1\{(\varepsilon_1,\varepsilon_2) \in A_{3 \cap 4,\beta,\gamma}(x)\} d\tilde F_{X,\varepsilon}(x,\varepsilon).
	\end{align}
	Under the assumption of independence of $(X, \varepsilon)$, we can rewrite (\ref{Q LRR prelim}) as:
	\begin{align}
		\label{ex_LRR}
		Q^\mathsf{LRR}(\beta,\gamma)  &= \frac{1}{4}\int P\left\{(\varepsilon_1,\varepsilon_2) \in A_{3 \cap 4,\beta,\gamma}(x)\right\} d\tilde F_X(x).   
	\end{align}
	We find the LRR by using this $Q^\mathsf{LRR}(\beta,\gamma)$ in (\ref{LRR3}).
	
	When $(\varepsilon_1,\varepsilon_2)$ fall in the region $A_{4,\beta,\gamma}(x)\cap A_{3,\beta,\gamma}(x)$, the game exhibits multiple equilibria. Our counterfactuals are most robust if this region is minimal. We cannot altogether disregard this region, because observations in the data may permit this region and it is possible they may arise in the counterfactual setting. Nevertheless, we can consider the value of the nuisance parameter(s) (in their identified set) that minimize the average region of multiple equilibria for the counterfactuals. (See Figure \ref{fig: entry game} for an illustration of this observation.)
	
	\subsubsection{A Switching Regression Model with Interval Outcomes \label{ex_interval}}
	Let us look for $Q^{\mathsf{LRR}}(\beta,\gamma)$ for the example from Section \ref{interval_data_ex}. We take $G$ to be a discrete distribution over $\{0,1\}$ and $\mu$ be the uniform distribution over $\{0,1\}$. Hence if we write
	\begin{align}
		\rho_{\beta_0,\gamma_0}(X,\gamma_0;\eta)  = g_1\left(\beta_0 D + X_1'\gamma_0 \right) \eta + g_2\left(\beta_0 D + X_1'\gamma_0 \right) (1 - \eta) + \varepsilon,
	\end{align}
	we find that
	\begin{align*}
		\Delta_{\beta_0,\gamma_0}(X,\varepsilon;\eta) &= \rho_{\beta_0,\gamma_0}(X,\varepsilon;\eta) - \int \rho_{\beta_0,\gamma_0}(X,\varepsilon;\eta) d\mu(\eta) \\
		&= \left(\eta  - \frac{1}{2}\right) \left(g_1\left(\beta_0 D + X_1'\gamma_0 \right) - g_2\left(\beta_0 D + X_1'\gamma_0 \right)\right).
	\end{align*}
	Therefore, given a counterfactual distribution $\tilde F(x,\varepsilon)$ of $(X,\varepsilon)$, 
	\begin{align}
		\label{bound4}
		Q^{\mathsf{LRR}}(\beta,\gamma) &= \int \int \Delta_{\beta,\gamma}^2(x,\varepsilon;\eta) d\mu(\eta) d\tilde F(x,\varepsilon) \\
		&= \frac{1}{4} \int \left(g_1\left(\beta d + x_1'\gamma \right) - g_2\left(\beta d + x_1'\gamma\right)\right)^2 d \tilde F_{X_1,D}(x_1,d),
	\end{align}
where $\tilde F_{X_1,D}$ denotes the joint distribution of $(X_1,D)$ under $\tilde F$. Thus, the $\kappa$-LRR of $\gamma_0$ can be rewritten as
	\begin{align}
		\label{LRR32}
		\Gamma_{\kappa}^{\mathsf{LRR}}(\beta) = \left\{\gamma \in \Gamma_P: Q^\mathsf{LRR}(\beta,\gamma)  \le Q^\mathsf{LRR}(\beta,\bar \gamma) + \kappa, \text{ for all } \bar \gamma \in \Gamma_P \right\},
	\end{align}
   where $\Gamma_P$ is as defined in (\ref{idset_interval}). Then we obtain the LRR set:
   \begin{align}
   	\label{B_LRR2}
   	\Theta_{\kappa,P}^{\mathsf{LRR}} = \left\{(\beta,\gamma): \gamma \in \Gamma_{\kappa}^{\mathsf{LRR}}(\beta), \beta \in B \right\}.
   \end{align}

   As mentioned previously, using this LRR set, one may construct the LRR set for other target parameters. Suppose that our ultimate interest is the Average Treatment Effect (ATE) denoted by $\psi(\beta_0,\gamma_0,\pi_0)$, where
   \begin{align}
   	\label{ATE}
   	\psi(\beta,\gamma,\pi) & = \mathbf{E}\left[ \alpha(X_1,1;\beta,\gamma,\pi) - \alpha(X_1,0;\beta,\gamma,\pi) \right],
   \end{align}
   where $\alpha(X_1,D;\beta,\gamma,\pi)$ is as defined in (\ref{alpha}). In this case, the LRR set for $\psi(\beta_0,\gamma_0,\pi_0)$ is given by
   \begin{align}
   	\Psi_{\kappa,P}^{\mathsf{LRR}} = \left\{\psi(\beta,\gamma,\pi): (\beta,\gamma) \in \Theta_{\kappa,P}^{\mathsf{LRR}}(\pi),\pi \in \Pi \right\},
   \end{align}
where $\Theta_{\kappa,P}^{\mathsf{LRR}}(\pi)$ is the same as $\Theta_{\kappa,P}^{\mathsf{LRR}}$, except that $\Theta_P$ is replaced by
\begin{equation}
	\label{idset_interval2}
	\Theta_P(\pi) = \left\{(\beta,\gamma): \mathbf{E}[\tilde Y_1 |X_1,D] \leq \alpha(X_1,D;\beta,\gamma,\pi) \leq \mathbf{E}[\tilde Y_2 |X_1, D], \text{a.e.} \right\}.
\end{equation}

	\section{Inference with the Locally Robust Refinement}
	\label{sec:Inf w LRR}
	
	\subsection{Inference from Moment Inequality Models\label{inference}}
	For the sake of concreteness, we focus on the case where the parameters are identified through moment inequality models and estimated using their sample moments.\footnote{Our LRR approach does not depend on a particular way in which partial identification of a parameter arises (i.e. whether by moment inequality models or alternative models). Hence, one can combine the LRR approach with other forms of partial identification by refining the identified set using the LRR criterion proposed in this paper.  For the sake of concreteness, we focus on the case where partial identification arises from moment inequality restrictions as this is the way partial identification often arises in practice.} Let us consider the following moment inequality models:
	\begin{align}
		\label{moment inequality}
		\mathbf{E}_P[m_j(Z_i;\beta_0,\gamma_0)] \le 0, \text{ for } j = 1,...,p,
	\end{align}
	for an observed i.i.d. random vector $Z_i$, where $m_j(\cdot;\beta_0,\gamma_0)$'s are moment functions. We assume that the moment restrictions define the identified set $\Theta_P$ so that for all $(\beta,\gamma) \in \Theta_P$, the moment restrictions are satisfied. 
	
	Define the sample analogue of the moment in (\ref{moment inequality}):
	\begin{align}
	\label{average moment}
	\overline m_j(\beta,\gamma) = \frac{1}{n}\sum_{i=1}^n m_j(Z_i;\beta,\gamma).
	\end{align}
	Let $\hat \sigma_j^2(\beta,\gamma)$ be the estimated variance of $m_j(Z_i;\beta,\gamma)$, i.e., 
	\begin{align}
	\hat \sigma_j^2(\beta,\gamma) = \frac{1}{n}\sum_{i = 1}^n \left(m_j(Z_i;\beta,\gamma) - \overline m_j(\beta,\gamma)\right)^2.
	\end{align}
	We let for $\kappa \geq 0$ (for instance, $\kappa = 0.01, 0.02$ or $0.03$)\footnote{The parameter $\kappa$ makes the inference slightly more conservative, which facilitates the development of the uniform inference (see the Supplemental Note). The choice of a positive constant $\kappa$ does not affect the uniform validity of the inference, as long as it is a fixed constant. In practice, values of $\kappa$ between 0.01 to 0.03 appear to work very well, as we show in the Monte Carlo simulations in Section 4 and the empirical application (Section 5).}
	\begin{align}
	\label{Q hat}
	\hat Q_{\kappa}(\beta,\gamma) = \sum_{j=1}^p\left[\frac{\overline m_j(\beta,\gamma)}{\hat \sigma_j(\beta,\gamma)}  - \kappa \right]_+,
	\end{align}
	where $[a]_+ = \max\{a,0\}$, $a \in \mathbf{R}$. When $\kappa = 0$, we simply write  $\hat Q(\beta,\gamma) = \hat Q_{0}(\beta,\gamma)$.
 Define
	\begin{align}
	\label{hat Gamma}
		\hat \Gamma_{\kappa}(\beta) =\left\{\gamma \in \Gamma: \hat Q_\kappa(\beta,\gamma) = 0 \right\}.
	\end{align}
   Then, we construct $T(\beta,\gamma)$ as
	\begin{align}
	\label{TLRR}
	T(\beta,\gamma) = \sqrt{n}\hat Q(\beta,\gamma).
	\end{align} 
	
	 As for critical values, we may consider two different approaches. The first approach is based on least favorable configurations. The second approach is a more refined version with enhanced power properties but with a higher computational cost. 
	 
	 As for the first approach, we consider the following:\footnote{We take this form of a test statistic for the sake of concreteness. One can use different functionals to produce different test statistics. See \cite{Andrews/Shi:13:Eca} for various functionals.}
	 \begin{align}
	 \label{hat Q star}
	 \hat Q^*(\beta,\gamma) = \sum_{j=1}^p\left[\frac{\overline m_j^*(\beta,\gamma) - \overline m_j(\beta,\gamma)}{\hat \sigma_j(\beta,\gamma)}\right]_+,
	 \end{align}
	 where
	 \begin{align}
	 \label{mj star}
	 \overline m_j^*(\beta,\gamma) = \frac{1}{n}\sum_{i=1}^n m_j(Z_i^*;\beta, \gamma),
	 \end{align}
	 and $Z_i^*$'s are resampled from the empirical distribution of $Z_i$'s with replacement. Let
	 \begin{align}
	 \label{hat T LRR*}
	 \hat T^*(\beta,\gamma) = \sqrt{n} \hat Q^*(\beta,\gamma).
	 \end{align}
	 
	 We take critical values $\hat c_{1-\alpha}(\beta,\gamma)$ to be the $1-\alpha$ quantile of the bootstrap distribution of $\hat T^*(\beta,\gamma)$. Then, we construct the confidence region for $(\beta_0,\gamma_0)$ as follows:
	 \begin{align}
	 \label{bootstrap CI0}
	 C_{1-\alpha}^\mathsf{LRR} = \left\{ (\beta,\gamma) \in \Theta: T(\beta,\gamma) \le \hat c_{1-\alpha}(\beta,\gamma), \text{ and } \gamma \in \hat \Gamma_{\kappa}^{\mathsf{LRR}}(\beta) \right\},
	 \end{align}
	 where
	 \begin{align}
	 \label{hat Gamma U}
	 \hat \Gamma_{\kappa}^{\mathsf{LRR}}(\beta) = \left\{\gamma \in \hat \Gamma_{\kappa}(\beta): \hat Q^\mathsf{LRR}(\beta,\gamma) \le \inf_{\bar \gamma \in \hat \Gamma_{-\kappa}(\beta)} \hat Q^\mathsf{LRR}(\beta,\bar \gamma) + 2 \kappa  \right\}
	 \end{align}
	 and $	\hat \Gamma_{\kappa}(\beta) $ is as defined in (\ref{hat Gamma}), and $\hat Q^\mathsf{LRR}(\beta, \gamma)$ denotes a consistent estimator of $Q^\mathsf{LRR}(\beta,\gamma)$ defined in (\ref{Q LRR}). (For example, in models of entry games or interval observations, this consistent estimator can be obtained by applying the sample analogue principle to the moments in (\ref{ex_LRR}) and (\ref{bound4}), once one parametrically specifies the distributions of $\varepsilon_1,\varepsilon_2,$ and $\varepsilon$.) We take the infimum of an empty set  to be infinity.  With $\gamma$ restricted to be in $\hat \Gamma_{\kappa}^{\mathsf{LRR}}(\beta)$, our confidence interval reflects the focus on the parameter values obeying the LRR hypothesis.
	
	As for the second approach, there are various alternative ways to construct inference that improves upon the one based on least favorable configurations. For example, one may use moment selection procedures as in \cite{Hansen:05:ET}, \cite{Andrews/Soares:10:Eca} and \cite{Andrews/Shi:13:Eca}, the contact set method in \cite{Linton/Song/Whang:10:JOE}, or a Bonferroni-based procedure as in \cite{Romano/Shaikh/Wolf:14:Eca}. Here we adapt the Bonferroni-based procedure to our context.
	
	First, we take $0< \alpha_1 < \alpha$ (say, $\alpha_1 = 0.005$) and find $\hat \kappa_{\alpha_1}(\beta,\gamma)$ using a bootstrap procedure. We define $\hat \kappa_{\alpha_1}(\beta,\gamma)$ as the $\alpha_1$ quantile of the bootstrap distribution of 
	\begin{align}
	\label{bootstrap}
	\min_{1 \le j \le k} \frac{\sqrt{n}(\overline m_j^*(\beta,\gamma) - \overline m_j(\beta,\gamma))}{\hat \sigma_j(\beta,\gamma)} ,
	\end{align}
	which uses $ \overline m_j^*(\beta,\gamma)$ defined in (\ref{mj star}). The uniform asymptotic validity of such a bootstrap procedure is well known in the literature. (See also Section \ref{subsubsec: general results} in the Supplemental Note for details.) Define
	\begin{align}
	\label{lambda}
	\hat \lambda_{j,\alpha_1}(\beta,\gamma) = \min \left\{\overline m_j(\beta,\gamma) - \frac{\hat \sigma_j(\beta,\gamma) \hat \kappa_{\alpha_1}(\beta,\gamma)}{\sqrt{n}},0\right\}.
	\end{align}
	Let
	\begin{align}
	\label{tilde Q*}
	\tilde Q^*(\beta,\gamma) = \sum_{j=1}^p\left[\frac{\overline m_j^*(\beta,\gamma) - \overline m_j(\beta,\gamma) + \hat \lambda_{j,\alpha_1}(\beta,\gamma)}{\hat \sigma_j(\beta,\gamma)}\right]_+.
	\end{align}
	We construct a bootstrap test statistic as follows:
	\begin{align}
	\label{tilde T LRR*}
	\tilde T^*(\beta,\gamma) = \sqrt{n} \tilde Q^*(\beta,\gamma).
	\end{align}
	We take critical values $\tilde c_{1-\alpha + \alpha_1}(\beta,\gamma)$ to be the $1-\alpha + \alpha_1$ quantile of the bootstrap distribution of $\tilde T^*(\beta,\gamma)$. Then, we construct the confidence region for $(\beta_0,\gamma_0)$ as follows:
	\begin{align}
	\label{bootstrap CI}
	\tilde C^{\mathsf{LRR}}_{1-\alpha} = \left\{ (\beta, \gamma)\in \Theta: T(\beta,\gamma) \le \tilde c_{1-\alpha + \alpha_1}(\beta,\gamma) \text{ and } \gamma \in \hat \Gamma_{\kappa}^{\mathsf{LRR}}(\beta) \right\}.
	\end{align}
    In the Supplemental Note, we establish the uniform asymptotic validity of the confidence region. In the next section, we describe how to implement this procedure for Example \ref{ex_interval} with both approaches to critical values.
    
    As emphasized in the introduction, the LRR-based confidence region $\tilde C^{\mathsf{LRR}}_{1-\alpha}$ should not be confused with one based on the identified set. This LRR-based confidence region does not reflect the empirical content of the model. Rather it shows the confidence region when the true parameter belongs to the LRR set, that is, the true parameter belongs to the set of parameter values at which the ASF is robust to the local perturbations of the equilibrium selection rules. One may use $\tilde C^{\mathsf{LRR}}_{1-\alpha}$ as a complement to the confidence region based on the identified set for $(\beta_0, \gamma_0)$ in empirical and policy analysis.
    
	\subsection{Monte Carlo Simulations}
	
	\subsubsection{Simulation Design\label{sec:design}}
	
	In this section, we present results from a Monte Carlo simulation study on the finite sample properties of the inference procedure developed previously.  We focus on the switching regression model example considered in Section \ref{ex_interval}, which will be the main specification of our empirical application introduced in the next section. 
	
	Let $\tilde Y_1, \tilde Y_2$ be as in Section \ref{interval_data_ex} and $\alpha(X_1,D; \beta_0,\gamma_0, \pi)$ as in (\ref{alpha}) with the following specification, also motivated by our empirical application:
	\begin{align}
		g_1(\beta_0 D + X_1'\gamma_0) &= \beta_0 D + X_1'\gamma_0, \text{ and } \\ \notag
		g_2(\beta_0 D + X_1'\gamma_0) &= \log\left(g_1(\beta_0 D + X_1'\gamma_0)\right).
	\end{align}
	We take $X_1 = 1$ for simplicity, we draw $\eta$ independently and with equal probability (i.e. $\pi_0 = 0.5$, implying an equal share of $\eta = 1$ and $\eta = 0$ in the population), and treatment is drawn independently for all $i$ with $D_i = 1$ with probability 0.3. The moment inequalities we have are as follows\footnote{
These inequalities can be generalized for a multidimensional $X_1$ by changing the indicator function on the left hand side to accommodate the different values of $X_1$.}: for $x_1 = 1$ and $d \in \{0,1\}$,
	\begin{align}
		\label{moment conditions}
		\mathbf{E}\left[(\tilde Y_1 - \alpha(1,d; \beta_0,\gamma_0, \pi_0)) 1\{D = d\}\right] &\le 0, \text{ and }\\
		\mathbf{E}\left[(\alpha(1,d;\beta_0,\gamma_0, \pi_0) - \tilde Y_2) 1\{D = d\}\right] &\le 0.
	\end{align}
	We consider two cases in our simulations: one where parameters $(\gamma_0, \beta_0, \pi_0)$ are such that the moments are close to equalities, and the other where they are not. For the first case, which we call Specification 1, we set $(\gamma_0,\beta_0, Z_1, Z_2) = (1,0.15,2.4,7)$. The choice of $\beta_0 = 0.15$ implies a return to treatment (for $\eta=1$) equal to 15\%. This specification corresponds to approximately 5\% of observations being top-coded.\footnote{As can be seen by the identified set (\ref{idset_interval}), setting one moment as an equality leaves the other moment as a strict inequality.  Therefore, we do not consider a ``least-favorable'' set-up with all moments holding as equalities.}  The second specification uses parameter values for $(\gamma_0, \beta_0, Z_1, Z_2)= (1,0.15,1.5,7)$ so that the moment inequalities are far from binding. This is done by keeping $(\gamma_0, \beta_0, Z_2)$ the same as the first specification, but decreasing $Z_1$ so that there are now approximately 10\% of observations top-coded, since we have an increased range of outcomes that are unobserved. In general, the identified set from the second specification will be larger than the first.
	
	For both cases, we show results for the identified set and for using LRR with and without the Bonferroni-based method of \cite{Romano/Shaikh/Wolf:14:Eca}.	For the LRR, we set $\kappa = 0.03$ throughout. The confidence interval follows the computation in (\ref{bootstrap CI}).  We set the nominal size at 0.05, the simulation number to 500, and the bootstrap number to 999 under a sample size of $n=500$. In this set-up,  our parameter space is such that $\beta_0 \geq 0$ and the ATE given in (\ref{ATE}) equals 0.145. %This could represent, for example, the effect of a job training program or of college, and would be useful for policies aimed at expanding such programs. 	
	
	\subsubsection{Results}
	
	The results across specifications and procedures are shown in Table \ref{MC_table}. They show the coverage of both the identified set and the LRR, as well as the length of the confidence sets. From the results, it is immediate that the LRR provides a smaller set of ATE $\psi(\beta_0,\gamma_0,\pi_0)$, implying less variation on predicted (counterfactual) outcomes $Y_i$. (The identified set, on the other hand, presents all possible values consistent with the data.) The coverage of the theoretical LRR is appropriate (at least 95\%) for values in the interior of the set, calculated using equations (\ref{bound4}) and (\ref{LRR32}).\footnote{In simulations available with the authors, the results improve as $n$ grows, as can be expected by our theoretical results. However, the computational cost also grows in these cases, particularly for the Bonferroni-based modification.} 
	
	\begin{center}
		[INSERT Table \ref{MC_table} HERE]
	\end{center}
	
	We also find that the Bonferroni-based method of \cite{Romano/Shaikh/Wolf:14:Eca} is effective at reducing the length of the confidence intervals, particularly for the identified set. It is most effective when the DGP is away from the configuration that is closest to moment equalities. However, this modification comes at the computational cost of calculating the term $\hat \kappa_{\alpha_1}$ via a bootstrap procedure for every value of $(\beta, \gamma)$ in a first step. This step could be time consuming when the dimension of $\gamma$ is large. 
	
	\subsection{Predictions Using the LRR and the Identified Set}
	
	We conduct two additional Monte Carlo exercises to provide additional insights on the benefits of using the Locally Robust Refinement. 
		
	Our first exercise compares the predictions for our parameter of interest ($\psi$) when using the LRR compared to using the identified set across Specifications 1 and 2. It illustrates when the LRR's predictions are closest to the true parameter values and when they are most informative (i.e. when the predictions vary the least).
	
\begin{figure}
\centering
\begin{subfigure}[t]{0.48\textwidth}
	\centering
	\includegraphics[width = \textwidth]{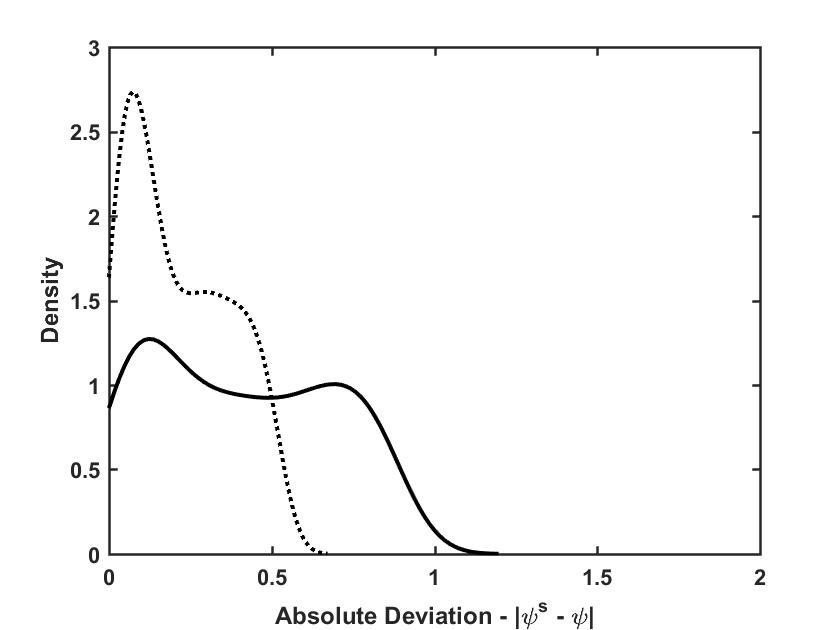}
	\caption{Specification 1}
\end{subfigure}%
~ 
\begin{subfigure}[t]{0.48\textwidth}
	\centering
	\includegraphics[width = \textwidth]{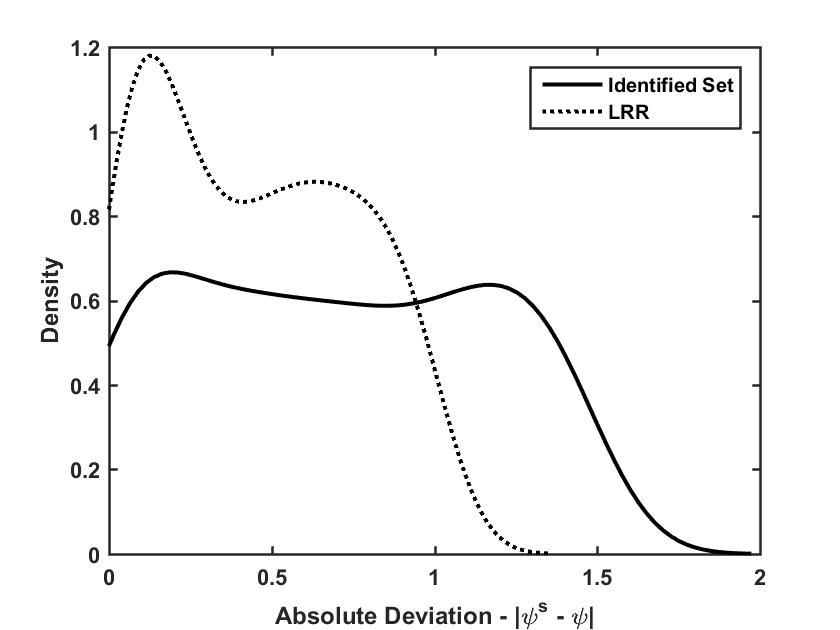}
	\caption{Specification 2}
\end{subfigure}

\caption{Distribution of the Absolute Difference Between Predicted ATEs and the True ATE}
\label{fig: MC MAD}

\medskip

\parbox{6.2in}{\footnotesize
	Notes: The figure presents the absolute deviation of the predicted statistic, $\psi^s$ drawn from the 95\% confidence set for the LRR or Identified set from the specifications in Section 4.2, relative to the parameter of interest, $\psi(\beta_0, \gamma_0,\pi_0)$, used for policy analysis. We present the uniformly-drawn parameters case, where we draw $\psi^s$ with equal probability from their respective sets $S=1000$ times. Drawing the furthest away parameter from the true values always yield the same parameter draw (so it is not pictured above). In the latter case, the mean absolute deviation is 0.521 for the LRR and 0.895 for the identified set for Specification 1, and 1.012 and 1.492 respectively for Specification 2. 
	\bigskip}
\end{figure}

	To do so, we draw $s=1,...,1000$ values $\psi^s$ from the average (conservative) 95\% confidence set for the LRR or the identified set described in Table \ref{MC_table}. These sets would be the predictions reported by an empirical researcher, who does not observe or point-identify the true $\psi(\beta_0, \gamma_0, \pi_0)$ due to top-coding. We draw these vectors in 2 different ways: (i) uniformly (i.e. any parameter within the appropriate set is drawn with equal probability) and (ii) furthest-case (i.e. we draw the parameters in each set that generate average predictions furthest away from the true values).\footnote{These sampling schemes represent different scenarios that illustrate the performance of our refinement. They may be thought of as different researcher priors on the true parameter. One can also perform the exercise under intermediate weighting schemes, obtaining similar conclusions.}  To showcase the results, in Figure \ref{fig: MC MAD}, we report the distribution of the absolute deviation $\left|\psi^s - \psi(\beta_0, \gamma_0,\pi_0)\right|$: i.e. how do the predictions using the LRR and the identified set differ from the true value.
	
\begin{figure}
\centering
\begin{subfigure}[t]{0.48\textwidth}
	\centering
	\includegraphics[width = \textwidth]{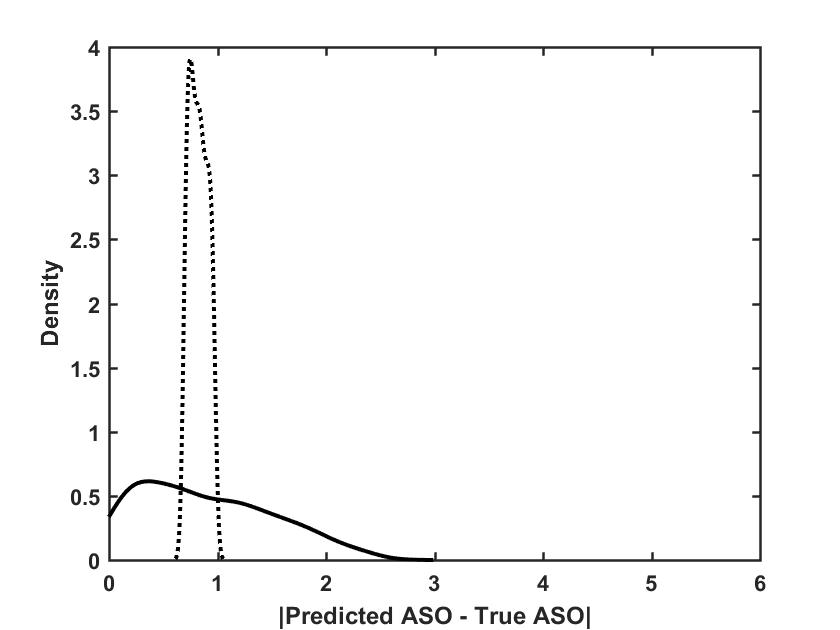}
	\caption{Average Prediction, under $\pi^s = 0.25$}
\end{subfigure}%
~ 
\begin{subfigure}[t]{0.48\textwidth}
	\centering
	\includegraphics[width = \textwidth]{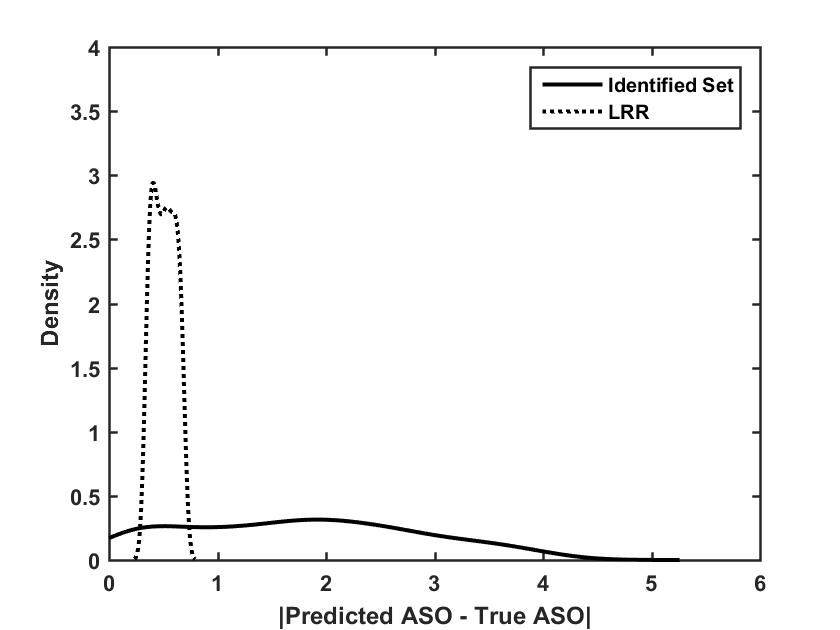}
	\caption{Average Prediction, under $\pi^s = 0.5$}
\end{subfigure}

\caption{Distribution of the Absolute Difference Between Predicted ASOs and the True ASO}
\label{no_LRRhyp}

\medskip

\parbox{6.2in}{\footnotesize
	Notes: The figure presents the absolute deviation of the predicted ASO, $\text{ASO}^s = \mathbf{E} \pi^s (\gamma^s + \beta^s D) + (1-\pi^s) \log(\gamma^s + \beta^s D)$, to the ASO under the true parameters $(\beta_0, \pi_0, \gamma_0) = (0.15, 0.5, 2.5)$ across draws $s=1,...,1000$. For the predicted outcome, we set $\pi^s$ at either 0.25 or 0.5 (true value), and draw the remaining parameters $S=1000$ times from the LRR or the identified set (calculated numerically). 
	\bigskip}
\end{figure}

	\begin{center}
		[INSERT FIGURES \ref{fig: MC MAD} AND \ref{no_LRRhyp} HERE]
	\end{center}
	
	As we can see, the LRR's predictions vary less than those with the identified set (they are often at least 30\% shorter), regardless of the specification or how the parameters are drawn. The use of the LRR seems particularly favorable when (i) the identified set is large (Specification 2), and (ii) when there is a significant chance of drawing extreme values of parameters from the identified set. This difference is most salient when the furthest-care scenario is particularly important to the researcher (for instance, when a bad policy could be very costly in terms of welfare).  Afterall, the LRR's objective is not about mean predictions, but rather as minimizing the worst case. 	
	
	The specifications above satisfy the LRR hypothesis, since the true $\psi(\beta_0, \gamma_0,\pi_0)$ is in the LRR set. Does the LRR still preserve these favorable properties even if the LRR hypothesis is not satisfied? We investigate this question in a last exercise. We set $(\beta_0, Z_1, Z_2)$ to the same values as Specification 1, but we increase $\gamma_0$ to 2.5. As a result, $\gamma_0$ is no longer in the LRR. We then compare the performance of the LRR and of the identified set in predicting the Average Structural Outcome (ASO) obtained using (\ref{g beta gamma Interval}).\footnote{In this setting, $\psi(\beta_0, \gamma_0,\pi_0)$ will generally satisfy the LRR hypothesis due to the parametrizations in this example. Hence as a target parameter that is not in the LRR set in this design, we consider the ASO in this exercise.} This outcome is unobserved, but it can be predicted based on the LRR and the identified set. The results are shown in Figure \ref{no_LRRhyp}. In contrast to Figure \ref{fig: MC MAD}, some draws of the identified set can outpredict the LRR. Afterall, the former contains the true parameters which are now absent in the LRR. Nevertheless, the LRR still generates more stable predictions, and its gains are still most noticeable and relevant when the identified set is large and/or worst-case outcomes are likely. Hence, while our results rely on the simulation designs motivated by our empirical application, the approach of LRR seems to generates predictions with stable performance regardless of whether the LRR hypothesis is satisfied or not.	
	
	\section{Empirical Illustration}
	
	We provide an empirical application to illustrate how our procedure works with data. The application considers the case of top-coded data, as in the Monte Carlo simulations. We focus on the returns to college education, a topic subject to a large body of work in economics (see \cite{Oreopoulos/Petronijevic:13:ERIC} for a survey). For the purpose of illustration, we choose a simplified approach and focus on the ATE defined in (\ref{ATE}).\footnote{One could extend our empirical analysis to incorporate potential endogeneity of the policy variable by using instrumental variables, and considering a Local Average Treatment Effect (LATE) which is the returns of college education on wages for the subgroup of individuals who are induced to attend college by a change in the instrument. Or alternatively, one could focus on the Marginal Treatment Effect or a Marginal Policy Relevant Treatment Effect as emphasized by \cite{Carneiro/Heckman/Vytlacil:11:AER}.} Depending on the question of interest, such a parameter could allow the evaluation of policies meant to increase enrollment, such as changing taxes or subsidies for college (see \cite{Dynarski/Scott:17:Tax} for an overview of different policies in place in the U.S. and their associated costs). These policies could have large fiscal and welfare consequences.
	
	In this set-up, the outcome ($Y_i$) is the log of the hourly wages, the policy treatment is attending college (i.e. $D_i \in \{0,1\}$, where $D_i = 1$ represents that $i$ attended college and $0$ otherwise) and $X_{1,i}$ are individual covariates such as gender and/or race. Finally, the decision to attend college may depend on unobserved types: some types may have higher wages and/or benefit more from college. We model this by the random variable $\eta$, which is assumed to be binary: the returns from treatment $D_i = 1$ may depend on whether unobserved type is high ($\eta = 1$) or low ($\eta = 0$) and this difference can exist throughout the distribution of $X_{1,i}$.\footnote{For illustration purposes, we assume $\eta$ is binary and independent of $X_1, D$. Both can be generalized, but that is beyond the scope of our exercise.} 
	
	Since top-coding is prevalent in wage data (it is present in such widely used datasets as the NLSY and the Current Population Survey - CPS) and such top-coding is not random (for example, college educated individuals are more likely to be top-coded since they have higher wages on average), this set-up is nested within Example \ref{ex_interval}. Hence, the identified set is given in (\ref{idset_interval}) and the LRR in (\ref{B_LRR}) can be used with the expressions defined in and above (\ref{LRR32}).
	
	We follow \cite{Romano/Shaikh:10:Eca} and use data from the 2000 Annual Demographic Supplement of the Current Population Survey (CPS), keeping individuals aged 20 to 24, white, with a primary source of income from wages and salaries and that worked at least 2 hours per week on average. Akin to their work, our covariates $X_{1,i}$ are a constant and a binary variable for Male/Female. We differ from their sample by keeping both college graduates and non-college graduates, since that is our counterfactual of interest. This yields a sample of 5816 individuals, of which 826 (14\%) are treated ($D_i = 1$). Details on the data construction are provided in the Supplemental Note.\footnote{\cite{Romano/Shaikh:10:Eca} took the empirical distribution of the data as a ``true'' probability distribution to perform Monte Carlo simulations. Here, as in typical empirical research, we regard the data as a sample from an unknown population.} We then explore the effects of different amounts of top-coding on the results.
	
	Following the large literature on Mincerian equations (see \cite{Heckman/Humphries/Veramendi:18:JPE} for a discussion), we use a linear specification for $g_1$:
	\begin{align*}
		g_1(\beta_0 D + X_1'\gamma_0) =\beta_0 D +  X_1'\gamma_0.
	\end{align*}
	For the reduced form for $\eta=0$, we parametrize $g_2(\beta_0 D + X_1'\gamma_0) = \log(g_1(\beta_0 D + X_1'\gamma_0))$.\footnote{While this is an illustration and other specifications could be used, this nonlinearity allows for increasing differentials across types, while its nonseparability allows for heterogeneous returns to college based on observable characteristics (see \cite{Barrow/Rouse:AERPP:05} for one such study).} We present confidence sets for both the identified set for the ATE, $\psi_0 = \psi(\beta_0,\gamma_0,\pi_0)$, and the LRR for $\psi_0$ using both conservative and a Bonferroni-correction based inference. We present results for this data under different top-coding amounts, imposing either 5\% or 10\% top-coding in the sample, under a value of top possible wages of $Z_2 = 10^8$.\footnote{We also set the upper bound on $\beta_0$ to 100\% (or 25\% per year), above all estimates typically found in the literature (see \cite{Oreopoulos/Petronijevic:13:ERIC}). This is a natural restriction using the literature's results, and helps speed up computation as we do not search for a priori unreasonable values of $\beta_0$).} The results are shown below. As standard, we report the returns to a year of college (i.e. $\psi_0/4$).
		\begin{center}
		[INSERT TABLE \ref{table emp} HERE]
	\end{center}
	Our estimates are largely consistent with the literature, including the ATE found in works as \cite{Carneiro/Heckman/Vytlacil:11:AER}, despite the different sample. While the identified set is large, including ATE close to 0 and others close to 25\%, the LRR is significantly shorter, between 3 and 11\% (for 5\% top-coding) and between 0 and 10\% (for 10\% top-coding). These results are also robust to different choices of $\kappa$.\footnote{We used $\kappa = 0.03$ which was shown to work well in the Monte Carlo simulations. When setting $\kappa = 0.01$, the LRR confidence sets are very similar (and shorter, since a larger $\kappa$ is more conservative): in the conservative inference case, they are $[0.061, 0.091]$ for 5\% top-coding and $[0, 0.091]$ for 10\% top-coding.} 
	
	To see the usefulness of the refinement in this context, consider one of the policies motivating this exercise. This could be a change in subsidies to college or expanding college loans with the purpose of a large increase in college attendance. In the presence of top-coding, the identified set cannot reject large rates of return for college (e.g. above 12\% a year) - suggesting a rationale for financially costly policies.  However, the LRR rejects these larger values, finding rates of return that are lower than 11\% and more compatible with those observed in the literature. The LRR suggests that more costly counterfactual policies should not be undertaken.  This rejection is not due to the informational content of the model, which is still captured by the identified set, but rather from the robustness of the counterfactual predictions. The larger returns to college found in the identified set seem to be driven by features of the model the researcher is less confident about.
	
	\section{Conclusion}
	
	This paper explores methods to deal with models with multiple reduced forms for counterfactual analysis. As mentioned by \cite{Eizenberg:14:ReStud}, there is an inherent tradeoff within the partial identification literature: while researchers may prefer not to impose strong identifying assumptions, they must confront the issue of counterfactuals with larger identified sets when doing policy analysis. 
	
	This paper points out that not all the values of the parameter are equally desirable for counterfactual analysis despite their empirical relevance. In particular, this paper focuses on the robustness property of the parameter values when the reduced form selection rule is locally perturbed. Thus the refinement looks at the subset of counterfactual scenarios that give most reliable predictions against a range of perturbations of the reduced forms (such as through changing equilibrium selection rules).
	
	There could be other ways to discriminate different values of the identified set depending on the purpose of the counterfactual analysis in the application. For example, one may consider local misspecification of causal relationships between certain variables. This is plausible when the researcher is not sure about the direction of causality between certain variables. Similarly as we do in this paper, one may want to find a refinement that is robust to this type of misspecification. A fruitful formulation of this approach is left for future research.
	
	\putbib[counterfactual]
\end{bibunit}

\section*{Tables and Figures}

\clearpage

\begin{table}[t]
\caption{\small The Empirical Coverage Probability and Average Length of Confidence
	Intervals for $\psi(\beta_{0},\gamma_0,\pi_0)$ at 95\% Nominal Level.}
\label{MC_table}
\small Coverage Probability

\begin{centering}
	\small
	\begin{tabular}{cc|cccc}
		\hline 
		\hline 
		& \multicolumn{1}{c}{} & \multicolumn{2}{c}{Specification 1} & \multicolumn{2}{c}{Specification 2}\tabularnewline
		\cline{3-6} 
		$\psi$ & \multicolumn{1}{c}{} & Least Favorable & RSW & Least Favorable &  RSW \tabularnewline
		\hline 
		Identified Set & $n=500$ & 1  & 0.996 & 1 & 1 \tabularnewline
	%	& $n=1000$   &  &   &  & \tabularnewline
		\hline 
		LRR & $n=500$ & 0.95 & 0.95  &  1 & 1  \tabularnewline
	%	& $n=1000$  &   &   &   &  \tabularnewline
		\hline 
		& \multicolumn{1}{c}{} &  &  &  &   \tabularnewline
	\end{tabular}
	\par\end{centering}

\small Average Length of CI

\begin{centering}
	\small
	\begin{tabular}{cc|cccc}
		\hline 
		\hline
		& \multicolumn{1}{c}{} & \multicolumn{2}{c}{Specification 1} & \multicolumn{2}{c}{Specification 2}\tabularnewline
		\cline{3-6} 
		$\psi$ & \multicolumn{1}{c}{} & Least Favorable & RSW & Least Favorable &  RSW \tabularnewline
		\hline 
		Identified Set & $n=500$ &     1.040 &  0.923   &     1.637  &  1.477  \tabularnewline
	%	& $n=1000$  &  &   &  &     \tabularnewline
		\hline 
		LRR & $n=500$ &      0.666 &    0.666   &  1.157 & 1.157  \tabularnewline
		%& $n=1000$  &   &   &   &  \tabularnewline
		\hline 
		& \multicolumn{1}{c}{} &  &  &  &   \tabularnewline
	\end{tabular}
	\par\end{centering}
\parbox{6.2in}{\footnotesize
	Notes: The first half of the table reports the empirical coverage
	probability of the confidence interval for $\psi$ computed with the bootstrap procedure described in the main text, and the second half reports its average length for the specifications described in the main text. The first inference procedure, denoted ``LF'', follows the conservative approach defined in the main text, while the second is denoted ``RSW'' and uses the Bonferroni-based modification of \cite{Romano/Shaikh/Wolf:14:Eca}.}
\end{table}

\begin{table}[t]
\caption{\small Confidence Intervals for $\psi$ (ATE), Parameter of Interest}

\label{table emp}

\small 

\begin{centering}
	\small
	\begin{tabular}{c|cc}
		\hline 
		\hline
		& 5\% top-coding & 10\% top-coding \\
		\hline 
		\\
		Identified Set - LF &  [0.023, 0.245] & [0, 0.245]  \tabularnewline 
		\\
		Identified Set - RSW &  [0.041, 0.245] & [0, 0.245]  \tabularnewline 
		\\
		\hline 
		\\
		LRR - LF  &  [0.034, 0.110] & [0, 0.091]  \\
		\\
		LRR - RSW  &  [0.041, 0.110] &  [0, 0.091]  \\
		\\
		\hline 
		\multicolumn{1}{c}{} &  &    \tabularnewline
	\end{tabular}
	\par\end{centering}
\parbox{6.2in}{\footnotesize
	Notes: The Table presents the results for the empirical application described in the main text, where we use the data and top-code it at different values (5\% or 10\%), assuming the top wage is $Z_2 = 10^8$. The first inference procedure, denoted ``LF'', follows the conservative approach defined in the main text - see equation (\ref{bootstrap CI0}), while the second is denoted ``RSW'' and uses the Bonferroni-based modification of \cite{Romano/Shaikh/Wolf:14:Eca}, see equation (\ref{bootstrap CI}).}
\end{table}

\FloatBarrier

\section{Appendix: Mathematical Proofs}

\subsection{Proof of Theorem \ref{thm: characterization}}

\noindent \textbf{Proof of Theorem \ref{thm: characterization}: } Let $U$ and $\mathcal{W}$ be the sets from which $\eta$ and $W = (X,\varepsilon)$ take values respectively. Let $\mathcal{H}$ be the collection of measurable maps $h: \mathcal{W} \times U \rightarrow \mathbf{R}$ such that 
\begin{align}
\label{center}
\int h(w,\eta)d\mu(\eta) = 0, \tilde F\text{-a.s.,} \text{ and } \int \int h^2(w,\eta)d\mu(\eta)d\tilde F(w) < \infty.
\end{align}
We endow $\mathcal{H}$ with the inner-product $\langle \cdot,\cdot \rangle$ as follows: for $h_1,h_2 \in \mathcal{H}$,
\begin{align}
\langle h_1,h_2 \rangle = \int \int h_1(w,\eta)h_2(w,\eta)d\mu(\eta)d\tilde F(w).
\end{align}
Then $(\mathcal{H},\langle \cdot,\cdot \rangle)$ is a Hilbert space (up to an equivalence class). Recall that all distributions $G \in \mathcal{G}$ are dominated by the uniform distribution $\mu$ over $H$. Define
$\| h \|^2 = \langle h, h \rangle$ and write
\begin{align}
\delta_\mathcal{G}(G',G) = \left\| \frac{dG'}{d\mu}  - \frac{dG}{d\mu} \right\|,
\end{align}
and $\textsf{ASO}_{\beta,\gamma}(G) = \int \int \rho_{\beta,\gamma}(w;\eta)dG(\eta|w)d\tilde F(w)$. If we define
\begin{align}
\Delta_{\beta,\gamma}(\eta,w) =  \rho_{\beta,\gamma}(w;\eta) - \int  \rho_{\beta,\gamma}(w;\eta) d\mu(\eta),
\end{align}
we can write
\begin{align*}
\sup_{G' \in \mathcal{G}: \delta_{\mathcal{G}}(G',G) \le K} \frac{|\textsf{ASO}_{\beta,\gamma}(G') - \textsf{ASO}_{\beta,\gamma}(G)|}{\delta_{\mathcal{G}}(G',G)} 
= \sup_{h' \in \mathcal{H}: \| h' \| = K}  \frac{\left\langle \rho_{\beta,\gamma}, h' \right\rangle}{\|h'\|} = \sup_{h' \in \mathcal{H}: \| h' \| = K}  \frac{\left\langle \Delta_{\beta,\gamma}, h' \right\rangle}{\|h'\|},
\end{align*}
where the first equality comes because $dG'/d\mu - dG/d\mu \in \mathcal{H}$ (from the fact that each density integrates to one) and the second equality comes from the fact that for all $h \in \mathcal{H}$, $\int h(w,\eta)d\mu(\eta) = 0$. Since $\Delta_{\beta,\gamma} \in \mathcal{H}$ by Assumption \ref{assump: L2 bound CASF}, using (\ref{center}) and Cauchy-Schwarz inequality, we can achieve the last supremum by taking $h'$ in the supremum to be $K \Delta_{\beta,\gamma}/\|\Delta_{\beta,\gamma}\|$. The result is nothing but $ \sqrt{Q^{\mathsf{LRR}}(\beta,\gamma)}$, completing the proof.\footnote{We note that the right hand side is not a function of $\delta$ or $K$. This is because the ASO is linear in $G$, and the dependence on those terms is removed by the difference operator defined on the left hand side.}  $\blacksquare$

\bigskip
\bigskip
\bigskip
\bigskip
\bigskip
\bigskip
\bigskip
\bigskip
\bigskip
\bigskip
\bigskip
\bigskip

\bigskip
\bigskip
\bigskip
\bigskip
\bigskip
\bigskip
\bigskip
\bigskip
\bigskip
\bigskip
\bigskip
\bigskip

\bigskip
\bigskip
\bigskip
\bigskip
\bigskip
\bigskip
\bigskip
\FloatBarrier
\clearpage
\pagebreak
\newpage
\appendix

\begin{bibunit}[econometrica] 
\vspace*{5ex minus 1ex}
\begin{center}
	\Large \textsc{Supplemental Note to ``Counterfactual Analysis under Partial Identification Using Locally Robust Refinement''}
	\bigskip
\end{center}

\date{%
	%TCIMACRO{\TeXButton{Today}{\today}}%
	%BeginExpansion
	\today%
	%EndExpansion
}

\vspace*{3ex minus 1ex}
\begin{center}
	Nathan Canen and Kyungchul Song\\
	\textit{University of Houston and University of British Columbia}\\
	\bigskip
	\bigskip
\end{center}

\makeatletter
\def\section{\@startsection{section}{1}
	\z@{.5\linespacing\@plus.5\linespacing}{.4\linespacing}{\large}}

\def\subsection{\@startsection{subsection}{2}
	\z@{.3\linespacing\@plus.3\linespacing}{-.5em}{\normalfont\bfseries}}
\makeatother

%\setcounter{section}{1}
%\setcounter{page}{1}

%\section*{A. Introduction}
%\setcounter{subsection}{0}
%\setcounter{equation}{0}

This note consists of three parts. In the first part (Appendix A), we present the proof of the uniform asymptotic validity of our inference, beginning with various auxiliary results. In the second part (Appendix B), we provide further details on the data used for the empirical application in Section 5. In Appendix C, we provide a notation table as a convenient reference. 

\section{Uniform Asymptotic Validity of the Confidence Region Under the LRR Hypothesis}
\subsection{Uniform Asymptotic Validity}

This section establish the uniform validity of the inference on LRR. To make explicit their dependence on $P$, we write $\tilde F$, $Q^{\mathsf{LRR}}(\beta,\gamma)$, and $\Gamma_{\kappa}^{\mathsf{LRR}}$ as $\tilde F_P$, $Q_P^{\mathsf{LRR}}(\beta,\gamma)$, and $\Gamma_{\kappa,P}^{\mathsf{LRR}}$  in this section. We first, make the following assumption.
\begin{assumption}
	\label{assump: uniform convergence}
	For each $\varepsilon>0$,
	\begin{align}
		\label{conv}
		\limsup_{n \rightarrow \infty} \sup_{P \in \mathbf{P}} P \left\{\sup_{(\beta,\gamma) \in \Theta}\left|\hat Q^{\mathsf{LRR}}(\beta,\gamma) - Q_P^{\mathsf{LRR}}(\beta,\gamma)\right| > \varepsilon\right\} = 0.
	\end{align}
\end{assumption}
Assumption \ref{assump: uniform convergence} requires uniform consistency of the sample criterion function for the LRR (uniform in $(\beta,\gamma) \in \Theta$ and $P \in \mathbf{P}$). Typically, $Q_P^{\mathsf{LRR}}(\beta,\gamma)$ takes the following form:
\begin{align*}
	Q_P^{\mathsf{LRR}}(\beta,\gamma) = \int q(x;\beta,\gamma) d\tilde F_P(x),
\end{align*}
for some map $q$ and a distribution $\tilde F_P$. Let $\hat F(x)$ and $\hat q(x;\beta,\gamma)$ be the estimated versions of $\tilde F_P(x)$ and $q(x;\beta,\gamma)$ using which we construct
\begin{align*}
	\hat Q^{\mathsf{LRR}}(\beta,\gamma) = \int \hat q(x;\beta,\gamma) d\hat F(x).
\end{align*}
(Note that when the researcher knows fully $\tilde F_P$ under the counterfactual scenario of focus, she can simply take $\hat F$ to be $\tilde F_P$.) Thus, we can find lower level conditions for (\ref{conv}) in terms of $\hat F(x)$ and $\hat q(x;\beta,\gamma)$ using standard arguments.

We let
\begin{align}
	\label{m_j,P}
	m_{j,P}(\beta,\gamma) = \mathbf{E}_P[m_j(Z_i;\beta,\gamma)], \text{ and }
	\sigma_{j,P}^2(\beta,\gamma) = \text{Var}_P(m_j(Z_i;\beta,\gamma)),
\end{align}
where $\text{Var}_P$ denotes the variance under $P$. Let
\begin{align}
	\mu_{i,j,P}(\beta,\gamma) = \frac{m_j(Z_i;\beta,\gamma)}{\sigma_{j,P}(\beta,\gamma)}.
\end{align}
Let us make the following assumption.
\begin{assumption}
	\label{assump: moment}
	(i) There exist $C>0$ and $q \ge 2$ such that
	\begin{align}
		\max_{1 \le j \le k} \sup_{P \in \mathbf{P}}\mathbf{E}_P\left[\sup_{(\beta,\gamma) \in \Theta}\left|\mu_{i,j,P}(\beta,\gamma)\right|^{q+1}\right] < C.
	\end{align}
	
	(ii) There exists $C>0$ such that for any $\delta >0$ and any $\theta  = (\beta,\gamma) \in \Theta$,
	\begin{align}
		\max_{1 \le j \le k}  \sup_{P \in \mathbf{P}}\mathbf{E}_P\left[\sup_{\theta' \in \Theta: \|\theta - \theta'\| \le \delta}\left|\mu_{i,j,P}(\theta) - \mu_{i,j,P}(\theta')\right|^2 \right] < C \delta^2.
	\end{align}
	
	(iii) $\Theta$ is compact.
\end{assumption}

Assumption \ref{assump: moment}(i) is a uniform moment condition. Assumption \ref{assump: moment}(ii) imposes the condition of $L_2$-equicontinuity on the function $\mu_{i,j,P}(\beta,\gamma)$. This condition, similar in spirit to \cite{Chen/Linton/vanKeilegom:71:Eca}, is used to verify the asymptotic equicontinuity of the empirical process involved in $\hat Q(\beta,\gamma)$ and its bootstrap version.

We now introduce an assumption on the limit Gaussian process of the empirical process. First, let us introduce some notation. Let $\{\nu_{j,P}(\beta,\gamma): (\beta,\gamma) \in \Theta\}$ be a Gaussian process having covariance function equal to $\text{Cov}_P(\mu_{i,j,P}(\beta,\gamma),\mu_{i,j,P}(\bar \beta,\bar \gamma))$, $(\beta,\gamma),(\bar \beta,\bar \gamma) \in \Theta$. Let for $M>0$,
\begin{align}
	\label{zeta M}
	\zeta_{M,P}^{\mathsf{sum}}(\beta,\gamma;r) &=  \sum_{j=1}^p \min \left\{[\nu_{j,P} + r_j]_+(\beta,\gamma), M \right\}, \text{ and }\\
	\zeta_P^{\mathsf{min}}(\beta,\gamma) &= \min_{1 \le j \le p} \nu_{j,P}(\beta,\gamma),
\end{align}
where $r = (r_j)_{j=1}^p$ is a vector of real measurable maps on $\Theta$. Define
\begin{align}
	R_M = \{(r_j)_{j=1}^p: r_j \text{ is measurable, and } -M \le r_j(\cdot) \le 0, \text{ for all } j=1,...,p\}. 
\end{align} 

\begin{assumption}
	\label{assump: nondeg gaussian}
	There exists $M_0>0$ such that the conditions below hold for $\zeta_{M,P}(\beta,\gamma;r) = \zeta_{M,P}^{\mathsf{sum}}(\beta,\gamma;r)$ or $\zeta_{M,P}(\beta,\gamma;r) = - \zeta_{P}^{\mathsf{min}}(\beta,\gamma)$.
	
	(i) For any $M > M_0$ and any $0< c < pM$,
	\begin{align}
		\sup_{P \in \mathbf{P}} \sup_{r \in R_M} \sup_{(\beta,\gamma) \in \Theta_P } P\{ c - \varepsilon \le \zeta_{M,P}(\beta,\gamma;r) \le c + \varepsilon\} \rightarrow 0,
	\end{align}
	as $\varepsilon \rightarrow 0$.
	
	(ii) For any $M > M_0$, there exists $\varepsilon_M>0$ such that for any $0< c < pM$, the slope of the CDF of $\zeta_{M,P}(\beta,\gamma;r)$ at $c$ is bounded from below by $\varepsilon_M$ for all $(\beta,\gamma) \in \Theta_P, r \in R_M$, and $P \in \mathbf{P}$.
\end{assumption}
Assumption \ref{assump: nondeg gaussian}(i) is a mild regularity condition for the Gaussian process $\nu_P$, which is generically satisfied because the Gaussian process has an unbounded support. (Note that $R_M$ is a collection of vectors of functions that are bounded between $-M$ and $0$.) The assumption can be violated when $\nu_{j,P}(\beta,\gamma)$ has zero variance at some point $(\beta,\gamma) \in \Theta_P$. Assumption \ref{assump: nondeg gaussian}(ii) is a technical condition which requires that $\zeta_{M,P}(\beta,\gamma;r)$ has CDF strictly increasing at any $c \in (0,p M)$ with slope bounded away from zero uniformly over $(\beta,\gamma) \in \Theta_P, r \in R_M$, and $P \in \mathbf{P}$. Note that when the Gaussian process $\nu_P$ has an unbounded support for all $(\beta,\gamma) \in \Theta_P$ and $P \in \mathbf{P}$, the support of $\zeta_{M,P}(\beta,\gamma;r)$ is $[0,pM]$ for all $(\beta,\gamma) \in \Theta_P, r \in R_M$, and $P \in \mathbf{P}$. Thus Assumption \ref{assump: nondeg gaussian}(ii) is not implausible.

The following theorem gives the uniform asymptotic validity of the confidence region $\tilde C_{1-\alpha}^{\mathsf{LRR}}$ defined in (\ref{bootstrap CI}).

\begin{theorem}
	\label{thm: uniform validity}
	Suppose that Assumptions \ref{assump: uniform convergence} and \ref{assump: moment} hold. Then,
	\begin{align}
		\liminf_{n \rightarrow \infty} \inf_{P \in \mathbf{P}} \inf_{(\beta,\gamma) \in \Theta_P} P \left\{ (\beta,\gamma) \in \tilde C_{1 - \alpha}^\mathsf{LRR}\right\} \ge 1 - \alpha.
	\end{align}
\end{theorem}

For the proof of uniform asymptotic validity, we approximate the bootstrap distribution by a truncated version of a functional of a Gaussian process. This facilitates the use of the continuous mapping theorem in \cite{Linton/Song/Whang:10:JOE} which does not require that the continuous functionals have a well defined limit.

The rest of Appendix A is devoted to proving Theorem \ref{thm: uniform validity}.

\subsection{Auxiliary Results I: General Results on Empirical Processes} \label{subsubsec: general results}
We introduce a theorem that gives the weak convergence of a stochastic process to a tight limit along a given sequence of probabilities. This result is based on Section 2.8.3 of \cite{vanderVaart/Wellner:96:WeakConvg}. We modify the results slightly because in our case the class of functions that index an empirical process depends on the underlying distribution $P \in \mathbf{P}$. The notation in this subsection is self-contained.

Let $\mathcal{T}$ be a given set and let $f_{j,P}(\cdot;\tau): \mathbf{R}^d \rightarrow \mathbf{R}$, $j=1,...,k$, be a function indexed by $\tau \in \mathcal{T}$, potentially depending on $P \in \mathbf{P}$. Define for each $j=1,...,k$,
\begin{align}
	\mathscr{F}_{j,P} = \{f_{j,P}(\cdot;\tau): \tau \in \mathcal{T}\}, \text{ and }  \mathscr{F}_P = \{f_P(\cdot;\tau): \tau \in \mathcal{T}\},
\end{align}
where $f_P(\cdot;\tau) = [f_{1,P}(\cdot;\tau),...,f_{k,P}(\cdot;\tau)]'$. Given i.i.d. random vectors $X_i$, we introduce a stochastic process on $\mathcal{T}$:
\begin{align*}
	\nu_{n,P}(\tau) = \frac{1}{\sqrt{n}}\sum_{i=1}^n (f_P(X_i;\tau) - \mathbf{E}_P[f_P(X_i;\tau)]).
\end{align*}
We let $\ell^\infty(\mathcal{T})$ be the set of bounded $\mathbf{R}^k$-valued functions on $\mathcal{T}$, and $\|\cdot\|_\infty$ denote the sup norm on $\ell^\infty(\mathcal{T})$, i.e., $\|\nu\|_\infty = \sup_{\tau \in \mathcal{T}} \|\nu(\tau)\|$, $\nu \in \ell^\infty(\mathcal{T})$, where $\|\cdot\|$ denotes the Euclidean norm in $\mathbf{R}^k$. We introduce a semimetric $\rho_P$ on $\mathcal{T}$ as follows:
\begin{align*}
	\rho_P(\tau_1,\tau_2) = \sqrt{\mathbf{E}_{P}\left[\|f_P(X_i;\tau_1) - f_P(X_i;\tau_2)\|^2 \right]}.
\end{align*}
Define the class of bounded Lipschitz functionals:
\begin{align}
	\label{BL}
	\quad \quad
	\text{BL}_1 = \left\{\rho: \ell^\infty(\mathcal{T}) \rightarrow \mathbf{R}: |\rho(\nu_1) - \rho(\nu_2)| \le \|\nu_1 - \nu_2\|_\infty, \sup_{\nu \in \ell^\infty(\mathcal{T})}|\rho(\nu)| \le 1 \right\}.
\end{align}
Write the $L_2(P)$ norm by $\|f\|_{P,2} = \sqrt{\mathbf{E}_P\|f(X_i)\|^2}.$ Given a class $\mathscr{F}_P$ of $\mathbf{R}^k$-valued measurable functions, let $N_{[]}(\varepsilon,\mathscr{F}_P,\|\cdot\|_{P,2})$ be the $\varepsilon$-bracketing entropy of $\mathscr{F}_P$ with respect to the $L_2(P)$ norm, which is the smallest number $N$ of the brackets $\{[\ell_j,u_j]\}_{j=1}^N$ such that each $f \in \mathscr{F}_P$ satisfies $\ell_j \le f \le u_j$ for some $[\ell_j,u_j]$ and $\|\ell_j - u_j\|_{P,2} \le \varepsilon$. (The inequalities required of the brackets are point-wise inequalities when $f$ is $\mathbf{R}^k$-valued.)

We also define $P^*$ and $\mathbf{E}_P^*$ to be the outer-probability and the outer-expectation under $P$. (See \cite{vanderVaart/Wellner:96:WeakConvg}, Section 1.2 for details.) The following theorem formulates conditions for the weak convergence of $\nu_{n,P}$ to a tight Gaussian limit uniformly over $P \in \mathbf{P}$.
\begin{theorem}
	\label{thm: weak conv}
	Suppose that the two conditions are satisfied.
	
	(i) For each $j=1,...,k$, there exists a measurable envelope $F_{j,P}$ for $\mathscr{F}_{j,P}$ such that
	\begin{align}
		\label{moment}
		\sup_{P \in \mathbf{P}}\mathbf{E}_P[F_{j,P}^2(X_i)] < \infty.
	\end{align} 
	
	(ii)
	\begin{align}
		\int_0^\infty \sup_{P \in \mathbf{P}} \sqrt{\log N_{[]}(\varepsilon\|F_{j,P}\|_{P,2},\mathscr{F}_{j,P},\|\cdot\|_{P,2})}d\varepsilon < \infty.
	\end{align}
	
	Then, we have
	\begin{align}
		\label{weak conv}
		\sup_{f \in \text{BL}_1} \sup_{P \in \mathbf{P}}\left|\mathbf{E}_{P}^*[f(\nu_{n,P})] - \mathbf{E}_{P}[f(\nu_{P})]\right| \rightarrow 0, \text{ as } n \rightarrow \infty, 
	\end{align}
	and
	\begin{align}
		\label{pre-Gaussianity}
		\lim_{\delta \downarrow 0}  \sup_{P \in \mathbf{P}} \mathbf{E}_P\left[ \sup_{\rho_P(\tau_1,\tau_2) < \delta} \|\nu_P(\tau_1) - \nu_P(\tau_2)\|\right] = 0,
	\end{align}
	where $\nu_P=[\nu_{1,P},...,\nu_{p,P}]'$ is an $\mathbf{R}^p$-valued Gaussian process with covariance kernel given by
	\begin{align*}
		\mathbf{E}[\nu_{j,P}(\tau_1)\nu_{\ell,P}(\tau_2)] = \text{Cov}_P\left(f_{j,P}(X_i;\tau_1),f_{\ell,P}(X_i;\tau_2)\right), \tau_1,\tau_2 \in \mathcal{T},
	\end{align*}
	for $j, \ell = 1,...,p$.
\end{theorem}

\noindent \textbf{Proof: } By following the proof of Theorem 2.8.4 of \cite{vanderVaart/Wellner:96:WeakConvg}, we find that Conditions (i) and (ii) give the asymptotic equicontinuity of the following form: for every $\varepsilon>0$,
\begin{align}
	\lim_{\delta \downarrow 0} \limsup_{n \rightarrow \infty} \sup_{P \in \mathbf{P}}P^*\left\{\sup_{\rho_P(\tau_1,\tau_2) < \delta} \left\|\nu_{n,P}(\tau_1) -  \nu_{n,P}(\tau_2)\right\| > \varepsilon \right\} = 0,
\end{align}
and $(\mathcal{T},\rho_P)$ is totally bounded uniformly over $P \in \mathbf{P}$ in the sense that for any $\varepsilon>0$, and any $P \in \mathbf{P}$ there exists an $m_\varepsilon$ number of $\varepsilon$-open balls (with respect to $\rho_P$) that cover $\mathcal{T}$ where $m_\varepsilon$ does not depend on $P$. Then both the results of (\ref{weak conv}) and (\ref{pre-Gaussianity}) follow by Theorem 2.6 of \cite{Gine/Zinn:91:AP}. $\blacksquare$
\medskip

We consider a bootstrap version of Theorem \ref{thm: weak conv}. Given a sample $\{X_1,...,X_n\}$, we let $\mathscr{G}_n$ be the $\sigma$-field generated by this sample. Let $\hat P$ be the empirical measure of $\{X_1,...,X_n\}$, and $\{X_1^*,...,X_n^*\}$ i.i.d. draws from $\hat P$. First, define
\begin{align*}
	\nu_{n,P}^*(\tau) = \frac{1}{\sqrt{n}}\sum_{i=1}^n (f_P(X_i^*;\tau) - \mathbf{E}_{\hat P}[f_P(X_i^*;\tau)]),
\end{align*}
which is the bootstrap counterpart of $\nu_{n,P}$.

\begin{theorem}
	\label{thm: boot weak conv}
	Suppose that the conditions of Theorem \ref{thm: weak conv} are satisfied. Then for any $\varepsilon>0$,
	\begin{align}
		\label{weak conv2}
		\sup_{P \in \mathbf{P}} P\left\{\sup_{f \in \text{BL}_1}\left|\mathbf{E}[f(\nu_{n,P}^*)|\mathscr{G}_n] - \mathbf{E}_{P}[f(\nu_{P})]\right| > \varepsilon \right\} \rightarrow 0,
	\end{align}
	as $n \rightarrow \infty$, where $\nu_P$ is the Gaussian process in Theorem \ref{thm: weak conv}.
\end{theorem}

\noindent \textbf{Proof: } The proof basically follows that of Lemma A.2 of \cite{Linton/Song/Whang:10:JOE}. The only difference here is that the class $\mathcal{F}_P$ depends on $P$. As seen in the proof of Theorem \ref{thm: weak conv}, $(\mathcal{T},\rho_P)$ is totally bounded uniformly over $P \in \mathbf{P}$. Then $(\mathcal{T}-\mathcal{T},\rho_P)$ is totally bounded uniformly over $P \in \mathbf{P}$ as well, where $\mathcal{T} -\mathcal{T}= \{\tau_1 -\tau_2: \tau_1,\tau_2 \in \mathcal{T}\}$. For any given $\varepsilon >0$, there is a map $\pi_\varepsilon: \mathcal{T} \rightarrow \mathcal{T}$ which takes $m_\varepsilon$ values and $\rho_P(\pi_\varepsilon \tau, \tau) \le \varepsilon$. Take any $f \in \text{BL}_1$, and consider
\begin{align}
	\left|\mathbf{E}[f(\nu_{n,P}^*)|\mathscr{G}_n] - \mathbf{E}_P f(\nu_P)  \right|
	&\le \left|\mathbf{E}[f(\nu_{n,P}^*)|\mathscr{G}_n] - \mathbf{E}[f(\nu_{n,P}^*\circ \pi_\varepsilon)|\mathscr{G}_n]  \right| \\ \notag
	& \quad +  \left|\mathbf{E}[f(\nu_{n,P}^*\circ \pi_\varepsilon)|\mathscr{G}_n] - \mathbf{E}_P f(\nu_P \circ \pi_\varepsilon)  \right| \\\notag
	& \quad +  \left|\mathbf{E}_P f(\nu_P \circ \pi_\varepsilon) - \mathbf{E}_P f(\nu_P)  \right|.
\end{align}
The last term vanishes uniformly in $f \in \text{BL}_1$  as $\varepsilon \rightarrow 0$ by (\ref{pre-Gaussianity}).  By using the proof of Lemma 2.1 in  \cite{Gine/Zinn:91:AP}, p.763, and the moment condition (\ref{moment}) to handle the remainder part of the process that is left over from truncation, we can show that the second term on the right hand side vanishes in probability uniformly over $P \in \mathbf{P}$ as $n \rightarrow \infty$. We deal with the first term using Le Cam's Poissonization lemma and following the proof of Theorem 2.2 of \cite{Gine:97:LectureNotes}, p.104.\footnote{The reference Theorem 2.2 of Gin\'{e} and Zinn (1991) on page 196 of \cite{Linton/Song/Whang:10:JOE} should be Theorem 2.2 of \cite{Gine:97:LectureNotes}, p.104.} More specifically, we define $\mathcal{T}(\varepsilon,\rho_P) = \{\tau_1 - \tau_2 \in \mathcal{T} - \mathcal{T}: \rho_P(\tau_1,\tau_2) < \varepsilon \}$, and note that
\begin{align}
	& \mathbf{E}_P\left[\mathbf{E}\left[ \sup_{\tau \in \mathcal{T}(\varepsilon,\rho_P)} \|\nu_{n,P}^*(\tau)\| |\mathscr{G}_n\right]\right] \\ \notag
	&\le \frac{e}{e-1}\left\{\mathbf{E}_P\left[\sup_{\tau \in \mathcal{T}(\varepsilon,\rho_P)} \left\|\frac{1}{\sqrt{n}}\sum_{i=1}^n(N_i -1)(f_P(X_i;\tau) - \mathbf{E}_P[f_P(X_i;\tau)]) \right\|\right]\right\}\\ \notag
	& \quad +  \mathbf{E}_P\left[\left|\frac{1}{\sqrt{n}}\sum_{i=1}^n(N_i -1)\right| \sup_{\tau \in \mathcal{T}(\varepsilon,\rho_P)} \left\|\frac{1}{n}\sum_{i=1}^n f_P(X_i;\tau) \right\|  \right],
\end{align}
where $N_i$ is i.i.d. Poisson random variables with mean $1$ independent of all the other random variables. We apply Theorem \ref{thm: weak conv} to the class $\mathcal{T}(\varepsilon,\rho_P)$, and find that both terms on the right hand, after taking supremum over $P \in \mathbf{P}$, vanish as $n \rightarrow \infty$ and $\varepsilon \rightarrow 0$. $\blacksquare$\medskip

The following lemma is the bootstrap version of Lemma A.1 of \cite{Linton/Song/Whang:10:JOE}. The proof of the lemma can be proceeded similarly as in the proof of Lemma A.1 of \cite{Linton/Song/Whang:10:JOE} using (\ref{weak conv2}) and is omitted.

\begin{lemma}
	\label{lemm: bootstrap CMT}
	Suppose that the bootstrap empirical process $\nu_{n,P}^*$ satisfies (\ref{weak conv2}) for any $\varepsilon>0$. Let $f_{n,P}$ be a continuous functional on $\ell^\infty(\mathcal{T})$ that is potentially $\mathscr{G}_n$-measurable such that for some possibly $\mathscr{G}_n$-measurable real-valued map $C_{n,P}$,
	\begin{align}
		|f_{n,P}(\nu_1) - f_{n,P}(\nu_2)| \le C_{n,P} \sup_{\tau \in \mathcal{T}} \left|\nu_1(\tau) - \nu_2(\tau)\right|,
	\end{align}
	for any $\nu_1,\nu_2 \in \ell_\infty(\mathcal{T})$, and for each $c \in A$ for some set $A \subset \mathbf{R}$,
	\begin{align}
		\lim_{\varepsilon \rightarrow 0} \limsup_{n \rightarrow \infty} \sup_{P \in \mathbf{P}} P\left\{|f_{n,P}(\nu_P) - c| \le C_{n,P} \varepsilon \right\} = 0.
	\end{align}
	Then, for each $c \in A$ and each $\varepsilon >0$,
	\begin{align}
		\limsup_{n \rightarrow \infty} \sup_{P \in \mathbf{P}} P \left\{ \left| P\left\{ f_{n,P}(\nu_{n,P}^*) \le c |\mathscr{G}_n \right\} - P\{f_{n,P}(\nu_P) \le c \}\right| > \varepsilon\right\} = 0.
	\end{align}
\end{lemma}\medskip

\subsection{Auxiliary Results II: Convergence of Stochastic Processes in Test Statistics}
We apply the general results on the empirical processes in the previous section to the convergence of stochastic processes in test statistics in this paper. For this subsection, we assume that Assumptions \ref{assump: uniform convergence}-\ref{assump: nondeg gaussian} hold.

For $j = 1,...,p$, let
\begin{align}
	\label{hat nu and tilde nu}
	\hat \nu_{n,j,P}(\beta,\gamma) &= \frac{\sqrt{n}(\overline m_j(\beta,\gamma) - m_{j,P}(\beta,\gamma))}{\hat \sigma_j(\beta,\gamma)}, \text{ and }\\ \notag
	\tilde \nu_{n,j,P}(\beta,\gamma) &= \frac{\sqrt{n}(\overline m_j(\beta,\gamma) - m_{j,P}(\beta,\gamma))}{\sigma_{j,P}(\beta,\gamma)},
\end{align}
and let $\hat \nu_{n,P} = [\hat \nu_{n,1,P},...,\hat \nu_{n,p,P}]'$, and $\tilde \nu_{n,P} = [\tilde \nu_{n,1,P},...,\tilde \nu_{n,p,P}]'$. Similarly, we also take for $j = 1,...,p$,
\begin{align}
	\label{hat nu and tilde nu bootstrap}
	\hat \nu_{n,j}^*(\beta,\gamma) &= \frac{\sqrt{n}(\overline m_j^*(\beta,\gamma) - \overline m_j(\beta,\gamma))}{\hat \sigma_j(\beta,\gamma)},\text{ and }\\ \notag
	\tilde \nu_{n,j}^*(\beta,\gamma) &= \frac{\sqrt{n}(\overline m_j^*(\beta,\gamma) - \overline m_j(\beta,\gamma))}{\sigma_{j,P}(\beta,\gamma)},
\end{align}
and let $\hat \nu_n^* = [\hat \nu_{n,1}^*,...,\hat \nu_{n,p}^*]'$ and $\tilde \nu_n^* = [\tilde \nu_{n,1}^*,...,\tilde \nu_{n,p}^*]'$. Let us first present the weak convergence results of the processes, $\tilde \nu_{n,j,P}(\beta,\gamma)$ and $\tilde \nu_{n,j}^*(\beta,\gamma)$. Let $\mathscr{G}_n$ be the $\sigma$-field generated by $\{Z_i\}_{i=1}^n$. Let $BL_1$ be the same as $BL_1$ in (\ref{BL}) except that $\mathcal{T}$ is replaced by $\Theta$.

\begin{lemma}
	\label{lemm: weak conv}
	There exists a tight Borel measurable Gaussian process $\nu_P$ on $\Theta$ such that
	\begin{align}
		\label{conv1}
		\sup_{f \in \text{BL}_1} \sup_{P \in \mathbf{P}} \left|\mathbf{E}_{P}^*[f(\tilde \nu_{n,P})] - \mathbf{E}_P[f( \nu_P)] \right| \rightarrow 0, \text{ as } n \rightarrow \infty,
	\end{align}
	and for each $\varepsilon>0$,
	\begin{align}
		\sup_{P \in \mathbf{P}}  P\left\{\sup_{f \in \text{BL}_1}\left|\mathbf{E}[f(\tilde \nu_n^*)|\mathscr{G}_n] - \mathbf{E}_P[f(\nu_P)] \right| > \varepsilon \right\}\rightarrow 0, \text{ as } n \rightarrow \infty.
	\end{align}
\end{lemma}

\noindent \textbf{Proof: } For the first statement, we use Theorem \ref{thm: weak conv}, and verify the conditions there. To map the current set-up to that of Theorem \ref{thm: weak conv}, we take $\mathcal{T} = \Theta$ and identify $\tau = (\beta,\gamma) \in \Theta$, $X_i = Z_i$, and $f_{j,P}(Z_i,\tau) = m_j(Z_i;\beta,\gamma)/\sigma_{j,P}(\beta,\gamma)$. Condition (i) of the theorem follows by Assumption \ref{assump: moment}(i), and Condition (ii) by Assumption \ref{assump: moment}(ii) and the assumption that $\Theta$ is compact. Thus (\ref{conv1}) follows by Theorem \ref{thm: weak conv}. By Theorem \ref{thm: boot weak conv}, the second statement follows similarly. $\blacksquare$

\begin{lemma}
	\label{lemm: consistency}
	For all $\varepsilon>0$,
	\begin{align*}
		\max_{1 \le j \le p} \sup_{P \in \mathbf{P}} P\left\{\sup_{(\beta,\gamma) \in \Theta} \left|\frac{\sigma_{j,P}(\beta,\gamma)}{\hat \sigma_j(\beta,\gamma)} - 1 \right| > \varepsilon \right\} \rightarrow 0,
	\end{align*}
	as $n \rightarrow \infty$.
\end{lemma}	

\noindent \textbf{Proof: } We can show this using Assumption \ref{assump: moment}(i) and (ii) and following the same steps in the proof of Lemma D.2 of \cite{Bugni/Canay/Shi:15:JOE}. Details are omitted. $\blacksquare$\medskip

Now combining Lemmas \ref{lemm: weak conv} and \ref{lemm: consistency}, we obtain the following result as a corollary.
\begin{lemma}
	\label{lemm: weak conv2}
	There exists a tight Borel measurable Gaussian process $\nu_P$ on $\Theta$ such that
	\begin{align}
		\label{conv12}
		\sup_{f \in \text{BL}_1} \sup_{P \in \mathbf{P}} \left|\mathbf{E}_{P}^*[f(\hat \nu_{n,P})] - \mathbf{E}_P[f( \nu_P)] \right| \rightarrow 0, \text{ as } n \rightarrow \infty,
	\end{align}
	and for each $\varepsilon>0$,
	\begin{align}
		\sup_{P \in \mathbf{P}}  P\left\{\sup_{f \in \text{BL}_1}\left|\mathbf{E}[f(\hat \nu_n^*)|\mathscr{G}_n] - \mathbf{E}_P[f(\nu_P)] \right| > \varepsilon \right\}\rightarrow 0, \text{ as } n \rightarrow \infty.
	\end{align}
\end{lemma} 

\noindent \textbf{Proof: } Note that
\begin{align*}
	\hat \nu_{n,P}(\beta,\gamma) = \tilde \nu_{n,P}(\beta,\gamma) + \tilde \nu_{n,P}(\beta,\gamma)\left( \frac{\sigma_{j,P}(\beta,\gamma)}{\hat \sigma_j(\beta,\gamma)} - 1 \right).
\end{align*}
By Lemma \ref{lemm: weak conv}, the process $\tilde \nu_{n,P}$ is asymptotically uniformly tight uniformly over $P \in \mathbf{P}$. By Lemma \ref{lemm: consistency}, the last term on the right hand side is $o_P(1)$ uniformly over $(\beta,\gamma) \in \Theta$ and over $P \in \mathbf{P}$. Lemma \ref{lemm: weak conv} then yields the first result of the lemma. The proof for the second result is similar. $\blacksquare$\medskip

\begin{lemma}
	\label{lemm: consist hat Q}
	\begin{align}
		\label{inc}
		\inf_{P \in \mathbf{P}} P\left\{ \Gamma_{\kappa,P}^\mathsf{LRR}(\beta) \subset \hat \Gamma_{\kappa}^\mathsf{LRR}(\beta), \forall \beta \in B \right\} \rightarrow 1,
	\end{align}
	as $n \rightarrow \infty$.
\end{lemma}	\medskip

\noindent \textbf{Proof: } In light of Assumption \ref{assump: uniform convergence}, it suffices to show that for all $\varepsilon>0$,
\begin{align*}
	\sup_{P \in \mathbf{P}} P\left\{\sup_{(\beta,\gamma) \in \Theta}|\hat Q(\beta,\gamma) - Q_P(\beta,\gamma)|> \varepsilon \right\} \rightarrow 0,
\end{align*}
as $n \rightarrow \infty$, where
\begin{align}
	\label{QP}
	Q_P(\beta,\gamma) = \sum_{j=1}^p \left[\frac{m_{j,P}(\beta,\gamma)}{\sigma_{j,P}(\beta,\gamma)}\right]_+.
\end{align}
First note that $\hat Q(\beta,\gamma)$ is a continuous map of $\overline m_j(\beta,\gamma)/\hat \sigma_j(\beta,\gamma)$. We use Assumption \ref{assump: moment}(ii) and the fact that $\Theta$ is compact to apply Glivenko-Cantelli lemma (i.e., the uniform law of large numbers) to the process $\overline m_j(\beta,\gamma)/ \sigma_{j,P}(\beta,\gamma)$. Then, the final result is obtained from Lemma \ref{lemm: consistency}. As the arguments are standard, details are omitted. $\blacksquare$

\subsection{Auxiliary Results III: Approximation by Truncated Functionals}
The test statistics in this paper are functionals of shifted empirical processes. As $n$ grows large, the shift can grow to infinity, which makes it cumbersome to apply the Continuous Mapping Theorem. To deal with this difficulty, we use the tightness of the limit Gaussian processes and approximate the functionals by truncated ones. 

We define
\begin{align}
	\label{q_{n,j,P}}
	q_{n,j,P}(\beta,\gamma) =\frac{\sqrt{n} m_{j,P}(\beta,\gamma)}{\sigma_{j,P}(\beta,\gamma)}, \text{ and }
	\hat q_{n,j,P}(\beta,\gamma) =\frac{\sqrt{n} m_{j,P}(\beta,\gamma)}{\hat \sigma_{j,P}(\beta,\gamma)}.
\end{align}
Given $\nu \in \ell^\infty(\Theta)$, let
\begin{align}
	\label{h defs}
	h_n(\nu)(\beta,\gamma) &= \sum_{j=1}^p[\nu_j + q_{n,j,P}]_{+}(\beta,\gamma),\\ \notag
	\hat h_n(\nu)(\beta,\gamma) &= \sum_{j=1}^p[\nu_j + \hat q_{n,j,P}]_{+}(\beta,\gamma),	
\end{align}
and for some constant $M >0$, define
\begin{align}
	\label{h defs2}
	\bar h_n^M(\nu)(\beta,\gamma) &= \sum_{j=1}^p\min\{[\nu_j + \max\{q_{n,j,P},-M\}]_{+}(\beta,\gamma),M\}, \text{ and }\\ \notag
	\hat h_n^M(\nu)(\beta,\gamma) &= \sum_{j=1}^p\min\{[\nu_j + \max\{\hat q_{n,j,P},-M\}]_{+}(\beta,\gamma),M\}.	
\end{align}
From here on, the constant $M>0$ is taken to be such that $M > M_0$ with the constant $M_0>0$ in Assumption \ref{assump: nondeg gaussian}. The following lemma delivers the approximation of functional $h_n$ by a truncated version.

\begin{lemma}
	\label{lemm: A}
	For any $\nu \in \ell^\infty(\Theta)$, and $n \ge 1$,
	\begin{align*}
		\sup_{(\beta,\gamma) \in \Theta_P} |h_n(\nu)(\beta,\gamma) - \bar h_n^M(\nu)(\beta,\gamma)| \le 3 M \sum_{j=1}^p A_{j,M}(\nu),
	\end{align*}
	where $A_{j,M}(\nu) = 1\left\{\sup_{(\beta,\gamma) \in \Theta} |\nu_j(\beta,\gamma)| \ge M \right\}$.
\end{lemma}

\noindent \textbf{Proof: } Let
\begin{align}
	h_n^M(\nu)(\beta,\gamma) &= \sum_{j=1}^p[\nu_j + \max\{q_{n,j,P},-M\}]_{+}(\beta,\gamma), \text{ and }\\
	\tilde h_n^M(\nu)(\beta,\gamma) &= \sum_{j=1}^p\min\{[\nu_j + q_{n,j,P}]_{+}(\beta,\gamma),M\}.
\end{align}
Let us also define
\begin{align}
	h_n^{M \Delta}(\nu)(\beta,\gamma) &=  |h_n(\nu)(\beta,\gamma) - h_n^M(\nu)(\beta,\gamma)|\\ \notag
	\bar h_n^{M \Delta}(\nu)(\beta,\gamma) &=  |h_n^M(\nu)(\beta,\gamma) - \bar h_n^M(\nu)(\beta,\gamma)|, \text{ and }\\ \notag
	\tilde h_n^{M \Delta}(\nu)(\beta,\gamma) &=  |\tilde h_n^M(\nu)(\beta,\gamma) - \bar h_n^M(\nu)(\beta,\gamma)|.
\end{align}
Since we can bound
\begin{align*}
	|h_n(\nu)(\beta,\gamma) - \bar h_n^M(\nu)(\beta,\gamma)| \le 
	h_n^{M \Delta}(\nu)(\beta,\gamma) + \bar h_n^{M \Delta}(\nu)(\beta,\gamma),
\end{align*}
it suffices to show that
\begin{align}
	\label{three stat}
	\sup_{(\beta,\gamma) \in \Theta_P} \tilde h_n^{M \Delta}(\nu)(\beta,\gamma) &\le M \sum_{j=1}^p A_{j,M}(\nu), \\ \notag
	\sup_{(\beta,\gamma) \in \Theta_P} \bar h_n^{M \Delta}(\nu)(\beta,\gamma) &\le M \sum_{j=1}^p A_{j,M}(\nu),\text{ and } \\ \notag
	\sup_{(\beta,\gamma) \in \Theta_P} |h_n^{M \Delta}(\nu)(\beta,\gamma) - \tilde h_n^{M \Delta}(\nu)(\beta,\gamma)| &\le M \sum_{j=1}^p A_{j,M}(\nu).
\end{align}

For each $j=1,...,p$ and $(\beta,\gamma) \in \Theta_P$, the absolute difference
\begin{align}
	|\min\{[\nu_j + q_{n,j,P}]_{+}(\beta,\gamma),M\} - \min\{[\nu_j + \max\{q_{n,j,P},-M\}]_{+}(\beta,\gamma),M\}|
\end{align}
is zero if $q_{n,j,P}(\beta,\gamma) \ge -M$. Suppose that $q_{n,j,P}(\beta,\gamma) < -M$. Then the difference is again zero if $\nu_j(\beta,\gamma) \le M$. Thus, we obtain the first statement in (\ref{three stat}). As for the second statement, note that for each $j=1,...,p$ and $(\beta,\gamma) \in \Theta_P$, the difference 
\begin{align}
	|[\nu_j + \max\{q_{n,j,P},-M\}]_{+}(\beta,\gamma) - \min\{[\nu_j + \max\{q_{n,j,P},-M\}]_{+}(\beta,\gamma),M\}|
\end{align}
is not zero only if $[\nu_j + \max\{q_{n,j,P},-M\}]_{+}(\beta,\gamma) \ge M$. Since $\max\{q_{n,j,P}(\beta,\gamma),-M\} \le 0$ for each $(\beta,\gamma) \in \Theta_P$, the latter inequality implies that $|\nu_j(\beta,\gamma)| \ge M$. Thus we obtain the second statement. As for the third statement in (\ref{three stat}), the difference between
\begin{align}
	&& |[\nu_j + q_{n,j,P}]_{+}(\beta,\gamma) - [\nu_j + \max\{q_{n,j,P},-M\}]_{+}(\beta,\gamma)| \text{ and }\\ \notag
	&& |\min\{[\nu_j + q_{n,j,P}]_{+}(\beta,\gamma),M\} - \min\{[\nu_j + \max\{q_{n,j,P},-M\}]_{+}(\beta,\gamma),M\}|,
\end{align}
is zero if $q_{n,j,P}(\beta,\gamma) > -M$. Assume that $q_{n,j,P}(\beta,\gamma) \le -M$. Then the last difference is not zero only if $[\nu_j + q_{n,j,P}]_{+}(\beta,\gamma) \ge M$ or
\begin{align}
	[\nu_j + \max\{q_{n,j,P},-M\}]_{+}(\beta,\gamma) \ge M.
\end{align}
The union of the latter two events, under $q_{n,j,P}(\beta,\gamma) \le -M$, is contained in the event $[\nu_j -M]_{+}(\beta,\gamma) \ge M$ which implies that $\nu_j(\beta,\gamma) > M$ or that $|\nu_j(\beta,\gamma)| > M$.
$\blacksquare$\medskip

\begin{lemma}
	\label{lemm: B}
	The following holds for each $\varepsilon,\eta>0$, as $n \rightarrow \infty$ and $M \rightarrow \infty$.
	
	\noindent (i) 
	\begin{align*}
		\sup_{P \in \mathbf{P}} \sup_{(\beta,\gamma) \in \Theta_P} P\left\{|\hat h_n(\hat \nu_{n,P})(\beta,\gamma) - \bar h_n^M(\hat \nu_{n,P})(\beta,\gamma)| > \varepsilon \right\} \rightarrow 0.
	\end{align*}
	
	\noindent (ii)
	\begin{align*}
		\sup_{P \in \mathbf{P}} \sup_{(\beta,\gamma) \in \Theta_P} P\left\{P\left\{|\hat h_n(\hat \nu_n^*)(\beta,\gamma) - \bar h_n^M(\hat \nu_n^*)(\beta,\gamma)| > \varepsilon |\mathscr{G}_n \right\} > \eta \right\} \rightarrow 0.
	\end{align*}
\end{lemma}

\noindent \textbf{Proof: } (i) First, we show that
\begin{align}
	\label{conv32}
	\sup_{P \in \mathbf{P}} \sup_{(\beta,\gamma) \in \Theta_P} P\left\{|h_n(\hat \nu_{n,P})(\beta,\gamma) - \bar h_n^M(\hat \nu_{n,P})(\beta,\gamma)| > \varepsilon \right\} \rightarrow 0.
\end{align}
By the first statement of Lemma \ref{lemm: weak conv2}, we find that
\begin{align*}
	\sup_{P \in \mathbf{P}} \left|P \left\{\sup_{(\beta,\gamma) \in \Theta}|\hat \nu_{n,j,P}(\beta,\gamma)|  \ge M \right\} - P \left\{\sup_{(\beta,\gamma) \in \Theta}|\nu_{j,P}(\beta,\gamma)|  \ge M \right\} \right| \rightarrow 0,
\end{align*}
where $\hat \nu_{n,j,P}(\beta,\gamma)$ denotes the $j$-th component of $\hat \nu_{n,P}(\beta,\gamma)$. By Lemma \ref{lemm: A}, it suffices to show that for any $j=1,...,p$,
\begin{align*}
	\sup_{P \in \mathbf{P}} M P \left\{\sup_{(\beta,\gamma) \in \Theta}|\nu_{j,P}(\beta,\gamma)|  \ge M \right\} \rightarrow 0,
\end{align*}
as $M \rightarrow \infty$. However, the last probability is bounded by
\begin{align}
	\label{bounds}
	M^{-1} \mathbf{E} \left[\sup_{(\beta,\gamma) \in \Theta}|\nu_{j,P}(\beta,\gamma)|^2 \right] &\le \frac{4}{M}  \left\| \sup_{(\beta,\gamma) \in \Theta}|\nu_{j,P}(\beta,\gamma)| \right\|_{\psi_2}^2 \\ \notag
	&\le \frac{C}{M},
\end{align}
for some $C>0$, where $\|\cdot\|_{\psi_2}$ denotes the Orlicz norm with $\psi_2(x) = \exp(x^2) - 1$. The first inequality follows because for any random variable $X$ and $ p \ge 1$, $\|X\|_{2p} \le p! \|X\|_{\psi_2}$. The last inequality follows from the proof of Corollary 2.2.8 of \cite{vanderVaart/Wellner:96:WeakConvg} and the bracketing entropy bound which comes from Assumption \ref{assump: moment} and the assumption that $\Theta$ is compact. The last term in (\ref{bounds}) vanishes as $M \rightarrow \infty$. Thus we obtain the convergence in (\ref{conv32}). Then the desired result follows from this convergence and Lemma \ref{lemm: consistency}.

(ii) The second statement follows similarly by using the second statement of Lemma \ref{lemm: weak conv2}, and following similar arguments in the proof of (i) using Lemma \ref{lemm: consistency}. $\blacksquare$

\begin{lemma}
	\label{lemm: nondeg}
	For each $M >M_0$ and each $c \in (0,pM)$,
	\begin{align}
		\lim_{\varepsilon \rightarrow 0} \limsup_{n \rightarrow \infty} \sup_{P \in \mathbf{P}} \sup_{(\beta,\gamma) \in \Theta_P} P\{|\bar h_n^M(\nu_{P})(\beta,\gamma) - c| \le \varepsilon \} = 0.
	\end{align}
\end{lemma}

\noindent \textbf{Proof: } Let $\tilde r_{j,P}(\beta,\gamma) = \max\{q_{n,j,P}(\beta,\gamma),-M\} 1\{(\beta,\gamma) \in \Theta_P\}$, $(\beta,\gamma) \in \Theta$. Then $\tilde r_{j,P} \in R_M$ for each $j=1,...,p$, and the desired result follows by Assumption \ref{assump: nondeg gaussian}. $\blacksquare$

\begin{lemma}
	\label{lemm: C}
	For each $M>M_0$ and $\eta \in (0,pM/2)$,
	\begin{align*}
		\quad \quad
		\sup_{P \in \mathbf{P}} \sup_{(\beta,\gamma) \in \Theta_P} \sup_{\eta \le c \le p M - \eta} \left|P\{\bar h_n^M(\hat \nu_{n,P})(\beta,\gamma) \le c \} - P\{\bar h_n^M(\nu_{P})(\beta,\gamma) \le c \} \right| \rightarrow 0,
	\end{align*}
	and, for each $\varepsilon>0$,
	\begin{align*}
		\sup_{P \in \mathbf{P}} \sup_{(\beta,\gamma) \in \Theta_P} P\left\{\sup_{\eta \le c \le p M - \eta} \left|P\left\{\bar h_n^M(\hat \nu_n^*)(\beta,\gamma) \le c|\mathscr{G}_n\right\} - P\left\{\bar h_n^M(\nu_P)(\beta,\gamma) \le c\right\} \right| > \varepsilon \right\}
		\rightarrow 0,
	\end{align*}
	as $n \rightarrow \infty$.
\end{lemma}

\noindent \textbf{Proof: } For each fixed $M>0$, the functional $\bar h_n^M$ is bounded Lipschitz with the Lipschitz coefficient equal to $p$. By Lemmas \ref{lemm: bootstrap CMT}, \ref{lemm: weak conv2}, and \ref{lemm: nondeg}, we have
\begin{align}
	\sup_{P \in \mathbf{P}} \sup_{(\beta,\gamma) \in \Theta_P} \left|P\{\bar h_n^M(\hat \nu_{n,P})(\beta,\gamma) \le c \} - P\{\bar h_n^M(\nu_{P})(\beta,\gamma) \le c \} \right| \rightarrow 0,
\end{align}
for each $c \in [\eta,pM - \eta]$. Due to Assumption \ref{assump: nondeg gaussian}(i), we can elevate this pointwise convergence with uniformity over $c \in [\eta,pM - \eta]$ by using the same arguments in the proof of Lemma 2.11 of \cite{vanderVaart:98:AsympStat}, p.12. Using Lemmas \ref{lemm: weak conv} and \ref{lemm: bootstrap CMT}, we can follow the same arguments to obtain the second statement for the bootstrap empirical process. $\blacksquare$\medskip

\begin{lemma}
	\label{lemm: unif tight}
	For any $\varepsilon>0$, there exists $M_\varepsilon>0$ such that
	\begin{align*}
		\limsup_{n \rightarrow \infty} \sup_{P \in \mathbf{P}} \sup_{(\beta,\gamma) \in \Theta_P} P\left\{h_n(\hat \nu_{n,P})(\beta,\gamma) \ge M_\varepsilon \right\} < \varepsilon,
	\end{align*}
	and
	\begin{align*}
		\limsup_{n \rightarrow \infty} \sup_{P \in \mathbf{P}} \sup_{(\beta,\gamma) \in \Theta_P} P\left\{P\left\{h_n(\hat \nu_n^*)(\beta,\gamma) \ge M_\varepsilon |\mathscr{G}_n\right\} > \varepsilon \right\} = 0.
	\end{align*}
\end{lemma}

\noindent \textbf{Proof:} Note that
\begin{align}
	\label{bound32}
	\sup_{(\beta,\gamma) \in \Theta_P} h_n(\nu)(\beta,\gamma) &= \sup_{(\beta,\gamma) \in \Theta_P} \sum_{j=1}^p[\nu_j + q_{n,j,P}]_{+}(\beta,\gamma) \\
	&\le \sup_{(\beta,\gamma) \in \Theta_P} \sum_{j=1}^p\left|\nu_j(\beta,\gamma) \right|, \notag
\end{align}
because $q_{n,j,P}(\beta,\gamma)\le 0$ for $(\beta,\gamma) \in \Theta_P$. By Lemma \ref{lemm: weak conv2}, $h_n(\hat \nu_{n,P})(\beta,\gamma)$ is uniformly tight uniformly over $P \in \mathbf{P}$, and thus we obain the first result.

As for the second result, note that
\begin{align*}
	& \sup_{P \in \mathbf{P}} \sup_{(\beta,\gamma) \in \Theta_P} P\left\{P\left\{h_n(\hat \nu_{n,P})(\beta,\gamma) \ge M_\varepsilon |\mathscr{G}_n\right\} > \varepsilon \right\}\\
	&\le \sup_{P \in \mathbf{P}} \sup_{(\beta,\gamma) \in \Theta_P} P\left\{|P\left\{h_n(\hat \nu_{n,P})(\beta,\gamma)  \ge M_\varepsilon |\mathscr{G}_n\right\} - P\left\{h_n(\nu_P)(\beta,\gamma)  \ge M_\varepsilon \right\}| > \varepsilon/2 \right\}\\
	& \quad +  \sup_{P \in \mathbf{P}} \sup_{(\beta,\gamma) \in \Theta_P} 1\left\{P\left\{h_n(\nu_P)(\beta,\gamma) \ge M_\varepsilon \right\} > \varepsilon/2 \right\}.
\end{align*}
The first supremum on the right hand side goes to zero as $n \rightarrow \infty$ by Lemma \ref{lemm: weak conv2}. By (\ref{bound32}) and the tightness of the Gaussian process $\nu_P$, we can choose a large $M_\varepsilon$ such that the last term becomes zero from some large $n$ on. $\blacksquare$

\begin{lemma}
	\label{lemm: D}
	For each $M>M_0$ and $\eta \in (0,pM/2)$,
	\begin{align*}
		\quad \quad
		\sup_{P \in \mathbf{P}} \sup_{(\beta,\gamma) \in \Theta_P} \sup_{\eta \le c \le p M - \eta} \left|P\{\hat h_n(\hat \nu_{n,P})(\beta,\gamma) \le c \} - P\{ h_n(\nu_{P})(\beta,\gamma) \le c \} \right| \rightarrow 0,
	\end{align*}
	and, for each $\varepsilon>0$,
	\begin{align*}
		\sup_{P \in \mathbf{P}} \sup_{(\beta,\gamma) \in \Theta_P} P\left\{\sup_{\eta \le c \le p M - \eta} \left|P\left\{\hat h_n(\hat \nu_n^*)(\beta,\gamma) \le c|\mathscr{G}_n\right\} - P\left\{h_n(\nu_P)(\beta,\gamma) \le c\right\} \right| > \varepsilon \right\}
		\rightarrow 0,
	\end{align*}
	as $n \rightarrow \infty$.
\end{lemma}

\noindent \textbf{Proof: } Since $\nu_P$ is a tight Gaussian process and $q_{n,j,P}(\beta,\gamma)\le 0$ for $(\beta,\gamma) \in \Theta_P$, for each $\varepsilon>0$, there exists $M_\varepsilon>0$ such that
\begin{align*}
	\sup_{P \in \mathbf{P}} \sup_{(\beta,\gamma) \in \Theta_P} P\left\{h_n(\nu_P)(\beta,\gamma) \ge M_\varepsilon \right\} < \varepsilon.
\end{align*}
The desired results follow from Lemmas \ref{lemm: B}, \ref{lemm: C} and \ref{lemm: unif tight}. $\blacksquare$

\subsection{Proof of Theorem \ref{thm: uniform validity}}
First note that we can write
\begin{align*}
	T(\beta,\gamma) = \hat h_n(\hat \nu_{n,P})(\beta,\gamma).
\end{align*}
Let us define
\begin{align}
	\label{def Ts}
	\bar T^{*}(\beta,\gamma) = \hat h_n(\hat \nu_n^*)(\beta,\gamma).
\end{align}
Note that from the definition of $\hat \lambda_{j,\alpha_1}(\beta,\gamma)$ in (\ref{lambda}),
\begin{align*}
	\frac{\sqrt{n}\hat \lambda_{j,\alpha_1}(\beta,\gamma)}{\hat \sigma_j(\beta,\gamma)} = \min \left\{\hat \nu_{j,n}(\beta,\gamma) + \hat q_{j,n,P}(\beta,\gamma) -  \hat \kappa_{\alpha_1}(\beta,\gamma),0\right\},
\end{align*}
where $\hat q_{n,j,P}$ is as defined in (\ref{q_{n,j,P}}). If we define
\begin{align*}
	\hat h_{\alpha_1, n}(\nu)(\beta,\gamma) = \sum_{j=1}^p[\nu_j(\beta,\gamma) + \min\{\hat \nu_{j,n}(\beta,\gamma) + \hat q_{n,j,P}(\beta,\gamma) - \hat \kappa_{\alpha_1}(\beta,\gamma),0\}]_{+},
\end{align*}
we can write
\begin{align*}
	\tilde T^{*}(\beta,\gamma) = \hat h_{\alpha_1, n}(\hat \nu_n^*)(\beta,\gamma).
\end{align*}
Let $\bar c_{1-\alpha+\alpha_1}(\beta,\gamma)$ be the $1-\alpha+\alpha_1$ quantile of the bootstrap distribution of $\bar T^{*}(\beta,\gamma)$, and recall that $\tilde c_{1-\alpha + \alpha_1}(\beta,\gamma)$ is the $1-\alpha+\alpha_1$ quantile of the bootstrap distribution of $ \tilde T^{*}(\beta,\gamma)$. We  compare $\bar c_{1-\alpha+\alpha_1}(\beta,\gamma)$ and $\tilde c_{1-\alpha + \alpha_1}(\beta,\gamma)$. For this, we first prepare the following lemma.
\begin{lemma}
	\label{lemm: conv kappa} 
	\begin{align*}
		\liminf_{n \rightarrow \infty} \inf_{P \in \mathbf{P}} \inf_{(\beta,\gamma) \in \Theta_P} P\left\{  \min_{1 \le j \le k} \hat \nu_{n,j,P}(\beta,\gamma) \ge \hat \kappa_{\alpha_1}(\beta,\gamma)\right\} \ge 1-\alpha_1,
	\end{align*}
	as $n \rightarrow \infty$.
\end{lemma}
\noindent \textbf{Proof: } By the second statement of Lemma \ref{lemm: weak conv2} and Assumption \ref{assump: nondeg gaussian}(ii), there exists $\kappa_{\alpha_1}(\beta,\gamma)$ such that for any $\varepsilon>0$,
\begin{align}
	\label{kappa}
	\sup_{P \in \mathbf{P}} \sup_{(\beta,\gamma) \in \Theta_P} P\left\{|\hat \kappa_{\alpha_1}(\beta,\gamma) - \kappa_{\alpha_1}(\beta,\gamma)| > \varepsilon \right\} = o(1),
\end{align}
as $n \rightarrow \infty$, and the density of $\min_{1 \le j \le k} \nu_{j,P}(\beta,\gamma)$ is bounded away from zero in a neighborhood of $\kappa_{\alpha_1}(\beta,\gamma)$. By using the definition of $\hat \kappa_{\alpha_1}(\beta,\gamma)$ and the second statement of Lemma \ref{lemm: weak conv2}, and the standard arguments of approximating an indicator function by a sequence of bounded Lipschitz functions (e.g., see the proof of Lemma A.1 of \cite{Linton/Song/Whang:10:JOE}), we have
\begin{align*}
	1 - \alpha_1 &\le \inf_{(\beta,\gamma) \in \Theta_P} P\left\{  \min_{1 \le j \le p} \hat \nu_{n,j}^*(\beta,\gamma) \ge \hat \kappa_{\alpha_1}(\beta,\gamma)|\mathscr{G}_n\right\}\\ \notag
	&\le \inf_{(\beta,\gamma) \in \Theta_P} P\left\{ \min_{1 \le j \le p} \nu_{j,P}(\beta,\gamma) \ge  \kappa_{\alpha_1}(\beta,\gamma)\right\} + o_P(1),
\end{align*}
uniformly over $P \in \mathbf{P}$ by Lemma \ref{lemm: consist hat Q}. By the first statement of Lemma \ref{lemm: weak conv2} and Lemma A.1 of \cite{Linton/Song/Whang:10:JOE}, the last infimum is equal to 
\begin{align*}
	&& \inf_{(\beta,\gamma) \in \Theta_P} P\left\{ \min_{1 \le j \le p} \hat \nu_{j,P}(\beta,\gamma) \ge  \kappa_{\alpha_1}(\beta,\gamma)\right\} + o(1)\\
	&\le \inf_{(\beta,\gamma) \in \Theta_P} P\left\{ \min_{1 \le j \le p} \hat \nu_{j,P}(\beta,\gamma) \ge  \hat \kappa_{\alpha_1}(\beta,\gamma)\right\} + o(1),
\end{align*}
uniformly over $P \in \mathbf{P}$, by Lemma \ref{lemm: consist hat Q} and (\ref{kappa}). $\blacksquare$\medskip

The lemma below compares the bootstrap critical values $\bar c_{1-\alpha+\alpha_1}(\beta,\gamma)$ and $\tilde c_{1-\alpha+\alpha_1}(\beta,\gamma)$.
\begin{lemma}
	\label{lemm: critical value comp}
	Suppose that the conditions of Theorem \ref{thm: uniform validity} hold. Then,
	\begin{align}
		\label{conv44}
		\liminf_{n \rightarrow \infty} \inf_{P \in \mathbf{P}}  \inf_{(\beta,\gamma) \in \Theta_P} P\left\{\tilde c_{1-\alpha+\alpha_1}(\beta,\gamma) \ge \bar c_{1-\alpha+\alpha_1}(\beta,\gamma) \right\} \ge 1 - \alpha_1.
	\end{align}
\end{lemma}

\noindent \textbf{Proof: } Denote by $E_n(\beta,\gamma)$ the event that for all $t \ge 0$,
\begin{align*}
	P\left\{ \hat h_{\alpha_1,n}(\hat \nu_n^*)(\beta,\gamma) \le t |\mathscr{G}_n\right\}
	\le P\left\{ \hat h_n(\hat \nu_n^*)(\beta,\gamma) \le t |\mathscr{G}_n\right\},
\end{align*}
and denote by $A_n(\beta,\gamma)$ the event that
\begin{align*}
	\min_{1 \le j \le p} \hat \nu_{n,j,P}(\beta,\gamma) \ge \hat \kappa_{\alpha_1}(\beta,\gamma).
\end{align*}
Then, since $E_n(\beta,\gamma)$ contains $A_n(\beta,\gamma)$,
\begin{align*}
	\liminf_{n \rightarrow \infty} \inf_{P \in \mathbf{P}} \inf_{(\beta,\gamma) \in \Theta_P} P E_n(\beta,\gamma) 
	\ge \liminf_{n \rightarrow \infty} \inf_{P \in \mathbf{P}} \inf_{(\beta,\gamma) \in \Theta_P} P A_n(\beta,\gamma) \ge 1 - \alpha_1,
\end{align*}
by Lemma \ref{lemm: conv kappa}. This completes the proof because $PE_n(\beta,\gamma)$ is bounded by the probability in (\ref{inc}). $\blacksquare$ \medskip

\noindent \textbf{Proof of Theorem \ref{thm: uniform validity}:} 
Note that
\begin{align*}
	& \limsup_{n \rightarrow \infty}\sup_{P \in \mathbf{P}} \sup_{(\beta,\gamma) \in \Theta_P}P \left\{ T(\beta,\gamma) > \tilde c_{1-\alpha+\alpha_1}(\beta,\gamma) \right\}\\
	&\le \limsup_{n \rightarrow \infty}\sup_{P \in \mathbf{P}} \sup_{(\beta,\gamma) \in \Theta_P} P \left\{ T(\beta,\gamma) > \overline c_{1- \alpha + \alpha_1}(\beta,\gamma) \right\} + \alpha_1 \text{ (by Lemma \ref{lemm: critical value comp})}.
\end{align*}
Hence the proof is complete once we show that the last limsup is bounded by $\alpha - \alpha_1$. However, this latter result comes from Lemma \ref{lemm: D} and Assumption \ref{assump: nondeg gaussian}(ii). The uniform validity of the confidence interval follows from this result and Lemma \ref{lemm: consist hat Q}. $\blacksquare$
\bigskip
\bigskip
\bigskip
\bigskip
\bigskip
\bigskip
\bigskip
\bigskip
\pagebreak

\section{Additional Details on the Data for the Empirical Application}

Our empirical application follows \cite{Romano/Shaikh:10:Eca} in using data from the 2000 Annual Demographic Supplement of the Current Population Survey (CPS) to study the relationship between covariates and wages in the presence of top-coding.  The CPS contains information on both individual characteristics (e.g. race, education, gender) and wages (from the primary earning job as well as those from other sources), making it appropriate for our analysis. However, the exercise in \cite{Romano/Shaikh:10:Eca} only covered college graduates. By contrast, our interest in the effects of college requires a sample with both college graduates and non-graduates. As a result, we must expand their sample to cover the latter group. Here, we describe our steps in this regard.

First, we used the IPUMS (Integrated Public Use Microdata Series) version of the CPS data - see \cite{IPUMS}. While the dataset is the same, this version integrates multiple years of the data and clarifies the original dictionary. As in \cite{Romano/Shaikh:10:Eca} p.186, we keep individuals who are between 20 and 24 of age (using the variable \texttt{AGE} in the dataset), white (using the variable \texttt{RACE}), with a primary source of income given by wages and salaries (using the variable \texttt{SRCEARN}, which stands for ``Source of earnings from longest job'') and that worked at least 2 hours per week on average (using the variable \texttt{UHRSWORKLY}). For wages, we used the variable \texttt{INCWAGE} which measures the total wage and salary income. This variable can be decomposed into ``Earnings from longest job'' (variable \texttt{INCLONGJ} in the CPS)	and ``Earnings from other work included wage and salary earnings'' (variable \texttt{OINCWAGE}). From their code and data, it appears that \cite{Romano/Shaikh:10:Eca} use the sum of the latter two.	

To code college graduates and non-college graduates, we use the standard definition from the CPS - and, to the best of our understanding - the one used by \cite{Romano/Shaikh:10:Eca}. That is, we code college graduates as those who have an ``Associate's Degree, Academic Program'', a ``Bachelor's Degree'', a ``Master's Degree'', a ``Professional Degree'' or a ``Doctorate Degree''. Non-graduates are those with ``Associate's Degree, Occupational/Vocational Program'', ``Some College/No Degree'', or those whose highest education is a higher school degree (or the equivalent) or lower.  We dropped individuals with missing data in one of these variables. This results in a final sample of 5816 individuals, of which 826 are college graduates (14.2\% of the sample), and 4990 non-graduates (85.8\% of the sample). For our empirical application, we code the treatment variable $D_i$ as equal to 1 for those in the first group, and 0 for the latter.\footnote{We note that our resulting sample size for \textit{college graduates} is 826, which is much larger than the 305 in \cite{Romano/Shaikh:10:Eca}. This is the case even though this subsample imposes the same restrictions outlined in their work. Unfortunately, we could not find details in their published version or in their Supplementary Material, such as further sample restrictions, which could explain these differences. We then extensively investigated the dataset to find why such differences could exist. First, we found that all of their original (distinct) observations are present in our sample of 826. We conclude that our sample comprehensively covers theirs and that they must have imposed additional restrictions. We extensively probed different variables to check for the latter, but we could not find a clear restriction. For instance, we could obtain the same number of observations, 305, by further restricting our sample to the first respondents in the household (\texttt{PERNUM} equal to 1) and by dropping observations those whose households lack a CPS identifier across surveys (based on the variable \texttt{CPSID}). However, these restrictions dropped over 40\% of the observations in the original \cite{Romano/Shaikh:10:Eca} sample, so we did not pursue it. We repeated our exercises with the non-IPUMS version of the CPS, available on the NBER webpage, to the same conclusions.}

\newpage

\section{Notation Table}
\begin{table}[h!]
	\small
	\begin{center}
		\small
		\begin{tabular}{lll}
			\hline\hline
			Notation & Description & Place of Def. \\
			\hline 
			& & \\
			
			$B_P$ &: Identified set for $\beta$, the parameter of interest, given $P \in \mathbf{P}$ &\\ 
			
			$\beta_0$ &: The parameter of interest. &\\
			
			$C_{1-\alpha}^\mathsf{LRR}$ &: A bootstrap confidence region for $\beta_0$ under the LRR hypothesis. & \\
			& (based on the least favorable configuration) & (\ref{bootstrap CI0}) \\
			$\tilde C_{1-\alpha}^{\mathsf{LRR}}$ &: A bootstrap confidence region for $\beta_0$ under the LRR hypothesis. & \\
			& (based on the approach of \cite{Romano/Shaikh/Wolf:14:Eca}) & (\ref{bootstrap CI}) \\
			
			$\hat c_{1-\alpha}(\beta)$ &: The $1-\alpha$ quantile bootstrap critical value. & Above (\ref{bootstrap CI0}) \\
			
			$\tilde c_{1-\alpha + \alpha_1}(\beta)$ &: The $1-\alpha+\alpha_1$ quantile bootstrap critical value. & Above (\ref{bootstrap CI})\\
			
			$F$ &: The distribution of $W = (X,\varepsilon)$ in the DGP for the data. & \\
			
			$\tilde F$ &: The counterfactual distribution of $W = (X,\varepsilon)$. & \\
			
			$G(\cdot|w)$ &: The conditional distribution of $\eta$ given $W=w$, as the reduced-&\\ 
			&form selection rule. &\\
			
			$\mathscr{G}_n$ &: The $\sigma$-field generated by the observations $\{(Y_i,X_i')'\}_{i=1}^n$. & \\
			
			$\Gamma(\beta)$ &: Identified set for $\gamma_0$, given $\beta$ and $P \in \mathbf{P}$ & (\ref{Gamma P beta})\\
			
			$\Gamma_{\kappa}^{\mathsf{LRR}}(\beta)$ &: The $\kappa$-LRR set. & (\ref{LRR3})\\
			
			$\hat \Gamma_{\kappa}(\beta)$ &: The estimated version of $\Gamma_P(\beta)$. & (\ref{hat Gamma}) \\
			
			$\hat \Gamma_{\kappa}^\mathsf{LRR}(\beta)$ &: The estimated upper LRR set & (\ref{hat Gamma U})\\
			
			$\gamma_0$ &: The nuisance parameter. &\\
			
			$h_n, \bar h_n^M, \bar h_n^M$ &: Functionals used to define various statistics. & (\ref{h defs}), (\ref{h defs2})\\
			
			$m_{j,P}(\beta,\gamma)$ &: The expected moment function in the moment inequality model & (\ref{m_j,P})\\
			
			$\overline m_j(\beta,\gamma)$ &: The average moment function in the moment inequality model & (\ref{average moment})\\
			
			$\hat \nu_{n,j,P}(\beta,\gamma)$ &: Empirical processes normalized by $\hat \sigma_j(\beta,\gamma)$ & (\ref{hat nu and tilde nu}) \\
			
			$\tilde \nu_{n,j,P}(\beta,\gamma)$ &: Empirical processes normalized by $\sigma_{j,P}(\beta,\gamma)$ & (\ref{hat nu and tilde nu}) \\
			
			$\hat \nu_{n,j}^*(\beta,\gamma)$ &: Bootstrap empirical processes normalized by $\hat \sigma_j(\beta,\gamma)$ & (\ref{hat nu and tilde nu bootstrap}) \\
			
			$\tilde \nu_{n,j}^*(\beta,\gamma)$ &: Bootstrap empirical processes normalized by $\sigma_{j,P}(\beta,\gamma)$ & (\ref{hat nu and tilde nu bootstrap}) \\
			
			$Q_P(\beta,\gamma)$ &: The population obj. function used to set-identify $(\beta_0,\gamma_0)$ & (\ref{QP}) \\
			
			$\hat Q(\beta,\gamma)$ &: The sample obj. function from the moment inequality model. & (\ref{Q hat})\\
			
			$\hat Q^*(\beta,\gamma)$ &: The bootstrap sample objective function. &\\
			& (based on the least favorable configuration.) & (\ref{hat Q star}) \\
			
			$\tilde Q^*(\beta,\gamma)$ &: The bootstrap sample objective function. &\\
			& (based on the approach of \cite{Romano/Shaikh/Wolf:14:Eca}) & (\ref{tilde Q*}) \\
			
			$Q^\mathsf{LRR}(\beta,\gamma)$ &: The population criterion for LRR. & (\ref{Q LRR})\\
			
			$T(\beta,\gamma)$ &: The test statistic. & (\ref{TLRR}) \\
			
			$\tilde T^{*}(\beta,\gamma)$ &: The bootstrap test statistic. & \\
			
			& (based on the approach of \cite{Romano/Shaikh/Wolf:14:Eca}) & (\ref{tilde T LRR*}) \\
			$\Theta_P$ &: Identified set for $(\beta_0,\gamma_0)$ given $P \in \mathbf{P}$ & Section \ref{overview_section}\\
			
			$\Theta_P^{\mathsf{LRR}}$ &: The set of $(\beta_0,\gamma_0) \in \Theta_P$ under the LRR hypothesis.& (\ref{B_LRR})\\
			& & \\
			\hline\hline
		\end{tabular}
	\end{center}
\end{table}
	
	\putbib[counterfactual]
\end{bibunit}
\end{document}